\documentclass[letterpaper,twocolumn,10pt]{article}
\usepackage{usenix2019_v3}

\usepackage{cite}
\usepackage{amsmath,amssymb,amsfonts}
\usepackage{algorithmic}
\usepackage{array,multirow,graphicx}
\usepackage{textcomp}
\usepackage{listings}
\usepackage{xcolor}
\usepackage{caption}
\usepackage{soul}
\usepackage{listofitems}
\usepackage{adjustbox}
\usepackage{fontawesome5}
\usepackage{enumitem}
\usepackage{tikz}
\usepackage{pgfplots}
\usepackage{makecell}
\usepackage{pifont}
\usepackage{booktabs}
\usepackage{subcaption}
\usepackage{float}
\usepackage{color, colortbl}
\usepackage{nicematrix}
\usepackage{comment}
\usepackage{titlesec}
\usepackage{placeins}
\usepackage[normalem]{ulem}

\usepackage{url}

\usepackage{breakurl}
\usepackage{hyperref}
\hypersetup{colorlinks}

\definecolor{grey1}{gray}{0.91}
\def\code#1{\texttt{#1}}

\definecolor{dgray}{gray}{0.25}
\definecolor{strings}{RGB}{0,0,128}
\definecolor{backcolour}{rgb}{0.95,0.95,0.92}
\definecolor{keywords}{RGB}{127,0,85}
\definecolor{darkred}{RGB}{139,0,0}
\definecolor{darkyellow}{RGB}{204,204,0}
\definecolor{darkblue}{rgb}{0.0, 0.0, 0.55}
\definecolor{vividviolet}{rgb}{0.62, 0.0, 1.0}
\definecolor{fuchsia}{rgb}{1.0, 0.0, 1.0}
\definecolor{shockingpink}{rgb}{0.99, 0.06, 0.75}
\definecolor{darkBlue}{RGB}{0, 0, 205}
\definecolor{darkgreen}{RGB}{0, 205, 0}
\definecolor{gray85}{gray}{0.85}
\definecolor{black}{gray}{0.05}
\newcommand\codenums{37 }

\usepackage[most]{tcolorbox} % Load the tcolorbox package
\definecolor{findingbg}{gray}{0.95} %light grey

\newcounter{findingcounter}
\newtcolorbox[auto counter]{finding}[1][]{
  colback=findingbg, % Background color of the box
  colframe=black, % Make frame color the same as the background
  left=0mm, % Left interior margin
  right=0mm, % Right interior margin
  top=0mm, % Top interior margin
  bottom=0mm, % Bottom interior margin
  boxsep=1mm, % Space between content and box
  arc=1.5mm, % No rounded corners
  outer arc=1.5mm, % No rounded corners on the exterior
  before upper={
  \refstepcounter{findingcounter}
  \textbf{\faSearch~Takeaway: }},
  #1
}

\usepackage{nopageno}

\begin{document}
%-------------------------------------------------------------------------------

%don't want date printed
\date{}

\title{\Large \bf A Mixed-Methods Study of Open-Source Software Maintainers On \\
Vulnerability Management and Platform Security Features}

% A Mixed-Methods Study Towards Understanding Why Previously Vulnerable Open-Source Projects Underuse Platform Security Features
% A Mixed-Methods Study Towards Understanding Challenges and Barriers Faced by Maintainers with Previously Vulnerable Open-Source Projects

%for single author (just remove % characters)
\author{
%{\rm Anonymous author(s)}
{\rm Jessy Ayala, Yu-Jye Tung, Joshua Garcia} \\ %}\\
University of California, Irvine
} % end author

\maketitle

\begin{abstract}
In open-source software (OSS), software vulnerabilities have significantly increased.
Although researchers have investigated the perspectives of vulnerability reporters and OSS contributor security practices, understanding the perspectives of OSS maintainers on vulnerability management and platform security features is currently understudied. 
In this paper, we investigate the perspectives of OSS maintainers who maintain projects listed in the GitHub Advisory Database.
We explore this area by conducting two studies: identifying aspects through a listing survey ($n_1=80$) and gathering insights from semi-structured interviews ($n_2=22$). 
Of the \codenums identified aspects, we find that supply chain mistrust and lack of automation for vulnerability management are the most challenging, and barriers to adopting platform security features include a lack of awareness and the perception that they are not necessary. 
Surprisingly, we find that despite being previously vulnerable, some maintainers still allow public vulnerability reporting, or ignore reports altogether. 
Based on our findings, we discuss implications for OSS platforms and how the research community can better support OSS vulnerability management efforts.
\end{abstract}

%In open-source software (OSS), software vulnerabilities have significantly increased. Although researchers have investigated the perspectives of vulnerability reporters and OSS contributor security practices, understanding the perspectives of OSS maintainers on vulnerability management and platform security features is currently understudied. In this paper, we investigate the perspectives of OSS maintainers who maintain projects listed in the GitHub Advisory Database. We explore this area by conducting two studies: identifying aspects through a listing survey (n1=80) and gathering insights from semi-structured interviews (n2=22). Of the 37 identified aspects, we find that supply chain mistrust and lack of automation for vulnerability management are the most challenging, and barriers to adopting platform security features include a lack of awareness and the perception that they are not necessary. Surprisingly, we find that despite being previously vulnerable, some maintainers still allow public vulnerability reporting, or ignore reports altogether. Based on our findings, we discuss implications for OSS platforms and how the research community can better support OSS vulnerability management efforts.

\section{Introduction}
The open-source software (OSS) ecosystem invites different kinds of reporters, contributors, and maintainers, creating a deeply integrated ecosystem crucial to the modern software supply chain.
OSS projects range from software on which many others depend on to software that depends on many others.
The pervasiveness of OSS is evident in a recent report by Synopsys~\cite{synopsys2024report}, where they found that 96\% of 1,067 codebases they scanned across 17 industries contained use of OSS. 

% the important role that maintainers play in either stopping vulns. at its track or letting it trickle down 
Given the pervasiveness of OSS, OSS software vulnerabilities deep within the software supply chain can have a crippling effect on society~\cite{heartbleed,shellshock,poodle}.
At the forefront of defending against these vulnerabilities are OSS maintainers who oversee those projects, where many may not have experience handling a security incident~\cite{wermke2022qual-OSSP}. 
OSS platforms such as GitHub offer various platform security features (e.g., dependency management~\cite{dpndbt}), but these features have been underutilized~\cite{ayala2023workflowssps,ayala2024glimpse,ayala2024secfeat,ayala2025investigatingvulnerabilitydisclosuresopensource}.
With the many responsibilities of being a maintainer, some have felt burned out~\cite{lawsonfeel} or left their project completely~\cite{nodeleave}.
The recent \code{xz-utils} incident~\cite{xzutils} in which an XZ Utils maintainer, who joined the project two years earlier, maliciously introduced a vulnerability in the project further highlights the challenges existing maintainers face in vulnerability management and mistrust in the software supply chain.
While several studies focus on the perspectives of vulnerability reporters \cite{akgul2023BBHperspectives, alexopoulos2021vuln-reporters, huang2016multiple, maillart2017eyes, ellis2022bounty, zhao2020control, li2022relational} and the security practices of OSS contributors \cite{hendrick2023openssf, wermke2022qual-OSSP}, there exist knowledge gaps concerning the perspectives of OSS maintainers, especially those who own previously vulnerable projects.

In this paper, we conduct a mixed-methods study that systematically identifies and quantifies the factors that affect OSS project maintainers’ involvement in conducting vulnerability management. 
Unlike prior work, we focus on recruiting OSS maintainers who maintain previously vulnerable projects, sourced from the GitHub Advisory Database and focusing on underused platform security features, and share experiences to allow researchers and OSS platforms to prioritize exploring efforts for improving the vulnerability management lifecycle, i.e., for those with little or no security background. To do so, we explore the following research questions:

\begin{center}
\noindent\fbox{%
    \parbox{0.971\columnwidth}{%
        \textbf{RQ1:}
        How do OSS maintainers with previously vulnerable OSS projects currently conduct vulnerability management? What are the challenges they face?
    }%
}
\end{center} 

\begin{center}
\noindent\fbox{%
    \parbox{0.971\columnwidth}{%
        \textbf{RQ2:}
        Why are platform security features underutilized in previously vulnerable OSS projects? What are the challenges and barriers to adopting such features?
    }%
}
\end{center} 

We conduct two studies, described in Section~\ref{sec:methods}, to address our RQs: a survey ($n_1=80$) to list current practices, general vulnerability management challenges, platform security feature challenges, platform security feature barriers, and wanted features or improvements, and conduct semi-structured interviews ($n_2=22$) study to contextualize overall survey results. Our primary contributions are as follows:

\vspace{-0.25cm}

\begin{itemize}
    \itemsep-0.4em 
    \item We are the first to investigate vulnerability management challenges that OSS maintainers, whose projects have a history of patched vulnerabilities, face regarding platform security features using GitHub Advisory Database.
    %\vspace{-0.1cm}
    \item We confirm previously discovered challenges, and further unveil novel challenges and barriers, OSS maintainers face when using current platform security features. Further, we systematically determine the breadth of security tooling and practices OSS maintainers use.
    %\vspace{-0.1cm}
    \item We provide an in-depth discussion, including suggestions and actionable items, which have corresponding insights to improve the OSS maintainer security processes, and have made our questionnaires for both studies publicly available~\cite{icse25repo} to facilitate future research.
\end{itemize}
%}

\vspace{-0.25cm}

Section~\ref{sec:parts} describes participant demographics and their backgrounds. We present results in Section~\ref{sec:results}. 
In summary, supply chain trust and a lack of understanding are the top general challenges. 
Limited automation, vulnerability scoring, and missing CI processes or features for vulnerability management, are particularly challenging and understudied for platform security features. 
The most prominent barriers to adopting platform security features are a lack of awareness, poor usability of such features, the complexity of their setup and usage, and the perception that they are unnecessary.
We also report wants from OSS maintainers to improve vulnerability management efforts, the most listed being assisted analysis and triaging, e.g., to automatically triage false positives, assisted platform security feature setup, e.g., setting up a security policy, and funding to be specifically used for security efforts, e.g., a bounty pool.

In Section~\ref{sec:discussion}, we discuss implications, i.e., based on significant challenges and barriers, for stakeholders involved in the vulnerability management lifecycle and future work that encourages researchers 
to develop tooling that minimizes regressions in the security context, advances current approaches working towards automated vulnerability scoring, and improving the usability of platform security features so that OSS maintainers have purpose to adopt them.
%, e.g., adding usability and purpose improvements usability and purpose in platform security features for OSS maintainers, 
%and e.g., large language models, 
%for further supporting OSS maintainers to conduct efficient and effective vulnerability analysis and triaging. 
Finally, we conclude in Section~\ref{sec:conclusion}.

%\vspace{-0.2cm}
\vspace{-0.1cm}

\section{Related Work}

% prior work grouping:
% - work focusing on vulnerability management in OSS ecosystem 
% - work focusing on OSS maintainers
% - work focusing on PSFs

%We discuss previous work through the following lenses:
%(1) current practices for vulnerability management in the OSS ecosystem,
%(2) the role that platform security features (PSFs) play in vulnerability management of the OSS ecosystem, and
%(3) challenges that OSS maintainers faced with respect to vulnerability management.

%bug bounty programs~\cite{finifter2013empirical,hata2017understanding}

% different perspectives of maintainers have been studied, their motivations and initial barriers.

% (1)
The openness of OSS invites various areas of studies, including automatic commit message generation~\cite{li2024only,cortes2014automatically,linares2015changescribe}, the pull-request development model~\cite{gousios2014exploratory,tsay2014influence,ford2019beyond,zhang2022pull},
and historical analysis of vulnerabilities~\cite{edwards2012historical,householder2020historical}.
To position our work in the body of literature on OSS, we focus our discussion of prior work through the following lenses:
(1) vulnerability management in the OSS ecosystem;
(2) the role that platform security features (PSFs) play in vulnerability management of the OSS ecosystem, including their benefits and challenges in adoption; and
(3) challenges and support for OSS maintainers.

%Both Wermke and Alomar highlight the importance of trust.

% disclosure 
% remediation
\vspace{0.1cm}
\noindent\textbf{Vulnerability management in the OSS ecosystem }
Our study centers around vulnerability management since it plays an important role in OSS. 
Lack of effective vulnerability management can affect trust~\cite{alomar2020you,fourne2023s}, causing maintainers and users to abandon a project. 
%The effectiveness of an OSS project's vulnerability management can either promptly fixed a vulnerability or cause it to trickle down the OSS ecosystem. Despite its importance, prior work has found vulnerability management practices to be scattered in OSS
%Prior work on vulnerability management has found that vulnerability fixes are often incompletes (i.e., requires additional commits to fix prior fixes)~\cite{bandara2020fix}, 
Prior work had found current OSS vulnerability management to be insufficient by mining repositories~\cite{ayala2023workflowssps,bandara2020fix,li2017large,ayala2024secfeat,ayala2024glimpse} and analyzing commits~\cite{bandara2020fix} and analyzing security patches~\cite{canfora2020process,li2017large}.   
Ayala et al.~\cite{ayala2023workflowssps} found a lack of security policy for many GitHub repositories.
Bandara et al.~\cite{bandara2020fix} and Li et al.~\cite{li2017large} found that initial security fixes are often insufficient, requiring multiple fixes.
Li et al. ~\cite{li2017large} 
%\josh{Cite even if it's the previous reference since this is a different sentence. Also add "et al.".} 
further found that a third of the security issues remain in repositories for three years before remediation, indicating a potential lack of effective vulnerability management practices. 

Prior work had also studied vulnerability management by interviewing security practitioners~\cite{alomar2020you} and OSS maintainers~\cite{wermke2022qual-OSSP}.
Alomar et al.~\cite{alomar2020you} studied vulnerability discovery and management processes from the perspective of organizations.
They 
%\josh{If this is the same authors from the previous sentence use "they".} 
found that organizations may take on vulnerability management for compliance reasons rather than improving the security of their products, with some organizations even asking pen-testers (external security experts) to lower the severity rating of vulnerabilities they discovered.   
Wermke et al.~\cite{wermke2022qual-OSSP} interviewed 27 OSS maintainers to understand their behind-the-scene processes (e.g., vulnerability management, processes for security and trust). Wermke found that most maintainers include a disclosure policy or contact for security issues in their projects.
However, security plays a minor role in the many hats that maintainers wear: Only a small number of maintainers (5/27) knew of a security role in their projects, four maintainers did not have a disclosure policy or contract for security issues, and few maintainers had dealt with security incidents.  
%Concerning security policies, they found most projects to contain at least some form of contact for security issues.
%In their pool of interviewees, only a few maintainers experienced security incidents.
\uline{Our work focuses on maintainers who have experienced security incidents and aims to understand how platforms like GitHub can better support vulnerability management and secure the OSS ecosystem.}

% (2)
% usefulness and existing issues of dependency management tools for vulnerability management 
%OSS platforms like GitHub offer various platform security features (PSFs) for vulnerability management. 
\vspace{0.1cm}
\noindent\textbf{The role of OSS platform security features (PSFs) in vulnerability management }
PSFs offered directly on the OSS platforms provide easy access for OSS maintainers to use for vulnerability management. 
PSFs such as dependency management tools (e.g., Dependabot~\cite{dpndbt}, Renovate Bot~\cite{renovatebot}, Greenkeeper~\cite{greenkeeper}) notify maintainers of dependency updates (e.g., vulnerability fixes, version updates).
%Dependabot ~\cite{mohayeji2023investigating,alfadel2021use,fischer2023effectiveness}, where most vulnerable dependencies identified by Dependabot were addressed within several days~\cite{mohayeji2023investigating} or a day~\cite{alfadel2021use}.  
Prior work~\cite{mohayeji2023investigating,alfadel2021use,fischer2023effectiveness} found Dependabot to be helpful in vulnerability management, where most vulnerable dependencies identified by Dependabot were addressed within several days~\cite{mohayeji2023investigating} or a day~\cite{alfadel2021use}.  
GitHub also found that 60\% more vulnerability-related, automated pull requests were merged in 2023 than in 2022~\cite{github2023report}.  
%Prior work has studied the use of dependency management tools, e.g., Dependabot, Renovate Bot, Greenkeeper, in OSS~\cite{mirhosseini2017can,he2023automating,wessel2021don,wessel2022bots,brown2019sorry}. 
%In particular, Dependabot has the highest visibility on Github since GitHub maintains it.
Mirhosseini et al.~\cite{mirhosseini2017can} found that maintainers update dependencies 1.6x more frequently with Greenkeeper than without because of the automated pull requests feature.
However, they also found that the automated pull request feature can create too many pull requests, generating significant noise and causing notification fatigue.
%The noise issue is not isolated to Greenkeeper; 
Others also found that the noise from dependency management tools to be a significant pain point for OSS maintainers despite their benefits~\cite{he2023automating,wessel2021don,brown2019sorry}.
Wessel et al.~\cite{wessel2022bots} introduced the idea of a meta-bot to mitigate noise.
He et al.~\cite{he2023automating} found that Renovate Bot is a popular migration target from Dependabot. 
One reason for this is Renovate Bot's auto-merge feature, which automatically merges pull requests for minor or patch dependency updates. 
% usefulness and existing issues of SCA tools for vulnerability management 
%Static code analysis tools are popularly used by maintainers to detect security issues.  
%Other popular PSFs OSS maintainers use are static code analysis tools for detecting security issues.

Aside from dependency management tools that retrospectively fix security issues,  
PSFs such as static code analysis tools can be used to proactively detect security issues; however, Ayala et al.~\cite{ayala2023workflowssps} found that they are underutilized by OSS maintainers.
One plausible explanation is the noise or false positives that static code analysis tools generate~\cite{wedyan2009effectiveness,zampetti2017open,johnson2013don}.
\uline{Our work confirms the aforementioned challenges of using PSFs, unveils further challenges, and provides suggestions to improve PSF adoption.}

%like dependency management tools, static code analysis tools are also plagued by significant noise or false positive results~\cite{wedyan2009effectiveness,zampetti2017open}.
%The amount of noise is a major reason for the under-utilization of static code analysis tools~\cite{johnson2013don}.

% (3)
% how donations and gamification help or not help with the vulnerability management efforts
Two popular forms of support for OSS maintainers prior work studied are gamification~\cite{moldon2021gamification,trockman2018adding,dabbish2012social} and donations~\cite{overney2020not,shimada2022github}.
%Prior work studied the use of gamification~\cite{moldon2021gamification,trockman2018adding,dabbish2012social} on GitHub.
Dabbish~\cite{dabbish2012social} found that maintainers appreciate \textit{``signals of attention,''} e.g., visible cues letting maintainers know that someone found their projects interesting. 
Trockman et al.~\cite{trockman2018adding} also found the use of gamification to be positive. In particular, badges such as quality assurance badges positively correlate with developers improving their test suites. 
They also speculate the usefulness of badges for bug fixing, stating, \textit{``One can imagine other badges with gamification value [e.g., around bug fixing] being used in the future to encourage desirable practices.''}
%Trockman~\cite{trockman2018adding} found that repository badges, e.g., badge status, test coverage, positively correlate with best practices. 
However, Molden et al.~\cite{moldon2021gamification} cautioned the use of gamification on GitHub, such as the daily activity streak features. 
%Molden et al.~\cite{moldon2021gamification} studied gamification in the context of GitHub's daily activity streak features.
They found that gamification may elicit unwanted behaviors, e.g., making contributions only to maintain activity streak.
%, while another study found that the most challenging aspect of bug bounty report review for OSS maintainers is how bug hunters are focused on money or CVEs, not software security~\cite{anon2025deepdive}.
%However, none studied gamification in the context of whether it benefits or harms vulnerability management efforts for OSS maintainers.
Besides gamification, another popular form of support is monetary incentives~\cite{overney2020not,shimada2022github}.  
Prior work found that GitHub Sponsors~\cite{shimada2022github}, a service provided by GitHub for maintainers to accept donations, are more effective than GitHub maintainers asking for donations using other donation platforms (e.g., PayPal, Patreon, Flattr)~\cite{overney2020not}.
%Overney et al.~\cite{overney2020not} found inconclusive evidence on the correlation between donations and the level of activity of OSS projects. 
\uline{Our work provides guidance on how gamification and funding can be used to promote PSF adoption and improve vulnerability management practices.}

% how toxic behaviors in OSS affect maintainers and their vulnerability management efforts
\vspace{0.1cm}
\noindent\textbf{Challenges and support for OSS maintainers }
Besides the challenges that afflict PSF usage, OSS maintainers face other challenges: toxicity~\cite{miller2022did,lawsonfeel,nodeleave,raman2020burnout,anon2025deepdive} and limited personpower~\cite{wermke2022qual-OSSP,lawsonfeel}. %, and attracting new contributors~\cite{wermke2022qual-OSSP,balali2020recommending,santos2022choose,lawsonfeel}.
Similar to other platforms such as Reddit, OSS platforms are also plagued by toxicity~\cite{miller2022did}.
Toxic security reporters, even with good intentions, may jeopardize the timeliness of maintainers fixing reported issues~\cite{destefanis2016software}.
Toxicity can strain maintainers emotionally~\cite{miller2022did}, causing maintainers to burn out~\cite{lawsonfeel} or even leave the project~\cite{nodeleave}.
In the context of dealing with bug bounty reports, the authors of ~\cite{anon2025deepdive} uncover that OSS maintainers find the diverted focus on money or CVEs from bug hunters and pressure to review reports the most challenging to deal with, especially with the threat of early vulnerability disclosures prior to patching. 
In addition to dealing with toxic reporters, which includes other maintainers~\cite{miller2022did}, OSS projects, particularly smaller ones, are generally limited in personpower.
%Prior work on attracting or retaining new contributors found a mismatch between the expectations of maintainers and contributors~\cite{santos2022choose} and 
On how to attract new contributors to increase personpower, Balali et al.~\cite{balali2020recommending} detailed 13 strategies.
However, most of the one-time contributors have no intention to become a long-time contributor~\cite{lee2017understanding}. 
%Our work highlights the challenges maintainers face in dealing with the unknowns (security issues) that come with the human-centric nature of OSS.
\uline{Unlike the aforementioned studies, our mixed-methods study focuses on the challenges arising for OSS maintainers with vulnerability management, especially platform security features involving the GitHub Advisory Database.}

\vspace{-0.1cm}

\section{Methodology}\label{sec:methods}

In this section, we provide an overview of our study approach and the outline of the semi-structured interviews.
We also discuss ethics, the qualitative coding process, report on our data collection, and discuss study limitations.
We designed and conducted two studies to investigate our research questions. 
A listing survey study to determine factors (collected from 04/2024 to 07/2024) and an interview study (conducted from 05/2024 to 08/2024).
The listing study allows us to understand current practices, challenges associated with, and needs of OSS project maintainers. We conduct a follow-up interview study to contextualize results.
\\

\vspace{-0.3cm}

%Our institutions’ review boards approved our study, and identifiable data was only available to authors listed on the corresponding study team tracking log. 

\noindent\textbf{Ethics } To comply with GitHub's terms of service~\cite{githubTOS}, we only reached out to maintainers who have publicly available contact information advertised as reachable to the general public, e.g., in text as a part of their profile introduction markdown, or through an external website, e.g., a personal homepage. 
Our institution’s ethics review board approved both studies. 
Participants signed consent forms detailing study plans and participant rights before data collection. Further, our study is GDPR-compliant.
We discuss ethics in recruiting specialized OSS maintainer populations in Section~\ref{sec:ethics}.

\vspace{-0.2cm}

\subsection{Listing survey study}\label{sec:lss}

To identify factors, e.g., challenges, that impact OSS vulnerability management practices, we conducted an online survey on OSS maintainers who own previously vulnerable projects using entries from the GitHub Advisory Database ($n_1=80$). 

\vspace{-0.2cm}

\subsubsection{Participant recruitment and piloting}\label{sec:recruit}

Before reaching out to OSS project maintainers who maintain projects with reviewed GitHub security advisories, we scrape the most recent 1,450 advisories per severity category, i.e., Low, Medium, High, and Critical, resulting in 5,096 advisories since there were under 1,450 Low reviewed advisories listed. 

We filtered the advisories that were tied to projects hosted on GitHub using advisory metadata, resulting in just over 2,000 unique GitHub projects. 
Two researchers then manually examined project organizations and profiles for metadata, discarding projects that do not fall under our IRB-approved protocol, which excludes subjects who reside in OFAC-sanctioned countries and regions, e.g., the \texttt{qq.com} email suffix.
Thus, leaving us with 1,920 unique GitHub projects to potentially recruit from.
To further comply with GitHub's terms of service~\cite{githubTOS}, two researchers then manually examined GitHub profiles, discarding subjects who do not have publicly available contact information advertised as reachable to the general public, through an external website, or social media. %Finally, leaving us with XXX unique GitHub projects that we recruited from.

We piloted the survey with ten respondents and reviewed the quality of responses, i.e., making sure the instructions were clear to create a list for each free-response question, for the listing survey study described in Section~\ref{sec:ldeets}. 
We then made wording updates to gather as many listing responses as possible to generate codes for categories shown in~\autoref{tab:bigasstable}.

\vspace{-0.2cm}

\subsubsection{Survey details}\label{sec:ldeets}

%We asked participants to list characteristics associated with conducting vulnerability management in four categories: benefits of bug bounty review, challenges of bug bounty review, helpful platform features, and wanted platform features. 
For each question, we informed that study participants should list all tooling and factors they use when conducting vulnerability management. 
We use an open-ended listing approach for free-response questions called free-listing~\cite{bernard2017research}, common in research when a domain is understudied, to elicit a full breadth of responses. 
Next, we asked participants to self-report OSS maintenance and industry experience, if they have a security background, project funding, and how often their vulnerability management process is reviewed. 
We ended with demographic questions to understand our population.
The full outline and questions asked can be found in our artifact~\cite{icse25repo}.

%\diffadd{
Prior work has not systematically determined the breadth of security tooling and practices OSS maintainers use. Instead, it has studied specific tooling in specific contexts, e.g., dependency tooling~\cite{alfadel2021dependabot,he2023automating,mirhosseini2017can}, static code analysis~\cite{johnson2013why,wedyan2009effectiveness,zampetti2017open}, and what OSS stakeholders implement for incident-handling~\cite{wermke2022qual-OSSP,ayala2023workflowssps}, while we focus on OSS maintainers since they are the primary point-of-contact for handling vulnerabilities. Prior work informs our survey design, which explores the breadth of approaches, challenges, and tooling GitHub OSS maintainers use to improve their projects’ security, which is currently understudied; our survey additionally draws from the \textit{Getting started} GitHub security features guide~\cite{ghsfeatures} and established initiatives that inform OSS maintainers on how to approach vulnerability management, e.g., guides from OpenSSF~\cite{openssfmg}.%}

\vspace{-0.2cm}

\subsubsection{Data analysis}

We analyzed listing survey responses with exploratory open-coding~\cite{saldana2021coding}.
Two researchers independently coded batches of ten responses at a time; further, resolving differences and updating the codebook after each batch.
Because the survey lent itself to multiple listings per category,
we were able to create an initial codebook from 27/80 (33.8\%) respondents, resulting in 23/\codenums (62.2\%) factors in~\autoref{tab:bigasstable}. In particular, two researchers analyzed the same 34 participant responses by engaging in open coding~\cite{charmaz2014constructing} and discussing the initial emerging themes, meeting five times. Both researchers then independently coded the remaining 46 responses in batches of 6 and met frequently to discuss findings and reach a consensus, resulting in 14 emerging codes after the initial analysis. The final 11 responses were received after the final codebook was established, %\diffadd{
i.e., coders met synchronously to fit them within the appropriate existing codes,%} 
containing \codenums codes in~\autoref{tab:bigasstable}.

\vspace{-0.3cm}

\subsection{Interview study}\label{sec:interviewstudy}

We asked consenting and interested listing study participants to participate in remote semi-structured interviews to learn more about why the identified factors and tooling listed are especially important to them and how such listings fit in current vulnerability management practices ($n_2 = 22$).

\vspace{-0.2cm}

\subsubsection{Participant recruitment and piloting}

Interviewees were a subset of the initial Listing survey study described in Section \ref{sec:lss}. 
If respondents were interested in participating in the interview study, they were directed to a Calendly \cite{calendly} space where they could join a publicly available Zoom link during specified time ranges and, if so, provide the GitHub project they oversee. 
%This also allowed us to access their corresponding non-identifiable demographic information and background as they were prompted to select if they were interested in participating in the interview study. 
%After reaching out using publicly listed contact information from GitHub projects, 25 responded, 22 of whom confirmed that they were available for an interview during the time ranges we advertised.
%\diffadd{
Of the 35/80 who responded \textit{Yes}, 19/35 scheduled a Calendly meeting. Using emails for recruitment, we contacted the other 16/35: 6 scheduled a meeting, and 10 did not respond. Of the 25 who scheduled a meeting, 3 were no-shows, resulting in 22 interviews.%}

We conducted pilots with three OSS project maintainers to test our semi-structured interview protocol, which can be found in our artifact~\cite{icse25repo}. 
We revised our interview questions based on the interviewees' quality of responses to narrow our research focus.
Details about participants regarding project information and interviews can also be found in our artifact~\cite{icse25repo}.

\vspace{-0.2cm}

\subsubsection{Interview details and structure}

We conducted semi-structured interviews with 22 OSS project maintainers, averaging 51 minutes and 59 seconds. Although survey participants were not compensated, interviewees were thanked with a virtual Visa \$20 gift card.

For reporting, we group the interview into six sections, each consisting of 1-2 opening questions, corresponding follow-up questions, and circling back to earlier ideas when applicable.
Before the main interview portion, we introduced our institution affiliations, an overall project overview, and our motivations. 
We went over how we only intended to keep audio, gathered consent for recording, and enabled closed captions and transcription.
We proceeded as follows:
%}
%We outline our semi-structured interview structure below
%and in our artifact~\cite{icse25repo}. 

\begin{comment}
\begin{figure}[ht]
  \includegraphics[width=\linewidth]{figs/interview_flow4.png}
  \caption{Illustration of the flow of topics of the semi-structured interview protocol. In each section, participants were presented with general questions and corresponding follow-ups but were free to diverge or skip questions entirely.}
  \label{fig:flow}
\end{figure}
\end{comment}

\vspace{0.05cm}
%\diffadd{
\noindent\textbf{0. Project maintainer duties } %The interview opens with a general question section about the project and our participant’s relation to it. 
This section is intended to ease nervous participants into the interview and establish initial context to later combine with actual repository data. 
%We report the demographics and combined data in~\autoref{tab:participants} and a more detailed breakdown of participants in~\autoref{sec:participants}.

\vspace{0.05cm}

\noindent\textbf{1. Current vulnerability management practices } This section explores
%investigates vulnerability management processes in the project. In particular, we were interested in 
structures that are not easily visible, such as who is involved in handling vulnerabilities, how vulnerabilities are \textit{actually} handled, and participant perceptions of why approaches are appropriate for them (Section~\ref{sec:currmanagement}). 
%We report these results in Section~\ref{sec:currmanagement}.

\vspace{0.05cm}

\noindent\textbf{2. Challenges with vulnerability management } This section explores
%investigates vulnerability management challenges in the project. In particular, we were interested in exploring 
past vulnerability management challenges encountered by our participants, e.g., through experiences, and how they have dealt with such challenges (Section~\ref{sec:genchallenges}).
%We report these results in Section~\ref{sec:genchallenges}.

\vspace{0.05cm}

\noindent\textbf{3. Challenges with platform security features } This section explores
%investigates platform security feature-specific challenges, i.e., challenges with tooling. We were interested in 
what overhead participants face with the PSFs they use and how such challenges affect their overall vulnerability management practices (Section~\ref{sec:psfchallenges}).
%We report these results in Section~\ref{sec:psfchallenges}.

\vspace{0.05cm}

\noindent\textbf{4. Barriers to adopting platform security features } This section explores 
%investigates platform security feature-specific barriers, i.e., what prevents the adoption of such tooling. We were interested in 
why participants refrain from using PSFs and their perceptions of what such features offer (Section~\ref{sec:psfbarriers}).
%We report these results in Section~\ref{sec:psfbarriers}.

\vspace{0.05cm}

\noindent\textbf{5. Opportunities for improvement and support } This section explores
%aims to elicit 
participants’ wants and needs for improving current PSFs and vulnerability management practices in general (Section~\ref{sec:psfwants}).%}
%We also ask for advice that would benefit other maintainers looking to make their project \textit{``secure''} (Section~\ref{sec:psfwants}).
%We report these results in Section~\ref{sec:psfwants}.

\vspace{-0.2cm}

\subsubsection{Data analysis}

All 22 audio-recorded interviews were transcribed and were checked for quality and accuracy by the researchers. 
Data was analyzed using thematic analysis~\cite{braun2006using}, starting with open coding~\cite{charmaz2014constructing} using each transcript, and developing thematic codes with axial coding~\cite{charmaz2014constructing} to describe common arising themes, e.g., \textit{Lack of awareness}. 
Two researchers then engaged in memo writing and constant comparison~\cite{glaser1965constant}, and inductive analysis based on grounded theory~\cite{charmaz2014constructing}. 
Both researchers analyzed the same 6 transcripts by engaging in open coding~\cite{charmaz2014constructing} and discussing the initial emerging themes, meeting four times.
During the analysis and iteratively discussing emerging themes across the 6 transcripts, both researchers engaged in axial coding~\cite{charmaz2014constructing}, analyzing the remaining 16 transcripts in parallel, and met frequently to discuss findings and reach consensus. 
Researchers were open to emerging themes; if a theme was adopted after discussion, researchers returned to previous interviews and re-coded accordingly.

\vspace{-0.2cm}

\subsection{Limitations}

Our methodology relies on self-reported data, which often entails substantial noise.
To mitigate this, we investigate our research questions using two studies, consisting of one survey study and one interview study; meanwhile, we maximize face validity by piloting each stage of the study, and revising procedures with feedback. 
Further, not all of our sample participants likely have extensive experience with OSS vulnerability management; however, we expect some participants with less OSS vulnerability management participation will have similar experiences or various perspectives.

While we had a sufficient number of participants to conduct our studies, i.e., 80 survey respondents and 22 interviewees, we cannot claim our results can be necessarily generalized to the OSS project maintainer demographic.
To mitigate this, we conduct two studies to ensure a holistic perspective, and our sample population varies in range regarding education level and years of involvement in managing OSS projects, as shown in~\autoref{tab:participants}.
Further, although the GitHub Advisory Database is our primary source of recruitment for the listing and interview studies, 1,920 unique GitHub projects are sampled and range in popularity, some with as many as 185 thousand stars and 400 thousand listed dependent GitHub projects.
\vspace{-0.2cm}

\section{Participants}\label{sec:parts}

~\autoref{tab:participants} summarizes participants’ self-reported demographics and experiences. 
We had 80 participants in the listing study and 22 interviewees. Participants were mainly from Europe and North America and were overwhelmingly male. 
This is consistent with other qualitative studies with OSS stakeholders as participants, e.g., ~\cite{liang2022understanding,huang2021fingerprints}.
Further, security experts from industry recommend organizations revisit their security policies and processes at least once a year~\cite{updatesec,updatesec2}, which is consistent with half of our study participants.

\begin{table}[th!]
    \begin{centering}
    \begin{adjustbox}{width=0.48\textwidth}
	\begin{tabular}{llrr}
            \toprule
            \midrule
            \centering
		& & \textbf{L} & \textbf{I} \\ 
		\midrule
            \midrule
		\textbf{Gender}     & Man & 71 & 19   \\
                                & Woman & 6 & 1   \\
                                & Non-binary & 3 & 2  \\
            \midrule
		\textbf{Age}        & 18-29 & 12 & 3  \\
                                & 30-39 & 28 & 8  \\
                                & 40-49 & 32 & 6  \\
                                & 50+ & 8 & 5  \\
		\midrule
		\textbf{Residence}  & Australia & 0 & 0  \\
                                & Africa & 1 & 0  \\
                                & Europe & 39 & 12  \\
                                & Middle East & 3 & 1 \\
                                & North America & 27 & 7  \\
                                & South America & 2 & 1  \\
                                & South Asia & 1 & 0  \\
                                & Southeast Asia & 1 & 0  \\
                                & Other & 6 & 1  \\
		\midrule
		\textbf{Education}  & $\le$ Completed high school & 3 & 1 \\
                                & Trade/technical/vocational & 1 & 1  \\
                                & College, no degree & 8 & 2  \\
                                & Associate’s degree & 1 & 1 \\
                                & Professional degree & 5 & 0  \\
                                & Bachelor's degree & 29 & 8  \\
                                & Graduate degree & 33 & 9 \\
            \midrule
		\textbf{Years working} & < 1 year & 4 & 1  \\
            \textbf{in industry} & 1-5 years & 10 & 2 \\
                                & 5-10 years & 4 & 1  \\
                                & 10+ years & 62 & 18 \\
                                %& None & 4 & 0 \\
            \midrule
		\textbf{Years in OSS} & < 1 year & 1 & 0  \\
            \textbf{maintainence} & 1-5 years & 16 & 4  \\
                                & 5-10 years & 18 & 5   \\
                                & 10+ years & 45 & 13  \\
            \midrule
		\textbf{Has a security} & Yes & 29 & 9  \\
            \textbf{background} & No & 51 & 13  \\
            \midrule
            \textbf{Vulnerabilty} & Every 1-3 months & 10 & 1  \\
		\textbf{management} & Every 6 months & 11 & 3  \\
            \textbf{process review} & Every year & 19 & 4  \\
            \textbf{or update} & Every 2+ years & 8 & 1  \\
            \textbf{frequency} & Never & 32 & 13  \\
            \midrule
		\textbf{OSS project has} & Yes & 31 & 7 \\
            \textbf{funding} & No & 49 & 15  \\
            \midrule
            \midrule
            \textbf{\# of participants} &  & \textbf{80} & \textbf{22}  \\
            \midrule
            \bottomrule
        \end{tabular}
    \end{adjustbox}
    \caption{The number of participants across the two studies along with their respective demographics, various backgrounds, and project details.  Acronyms listed represent, respectively: listing (\textbf{L}) survey study and interview (\textbf{I}) study.}
    \label{tab:participants}
    \end{centering}
\end{table}

In the listing survey and interview studies, our participants are OSS project maintainers who have received and patched vulnerabilities as seen in the GitHub Advisory Database~\cite{gad}. 
We are especially curious about this subset of OSS project maintainers because although they have experience triaging vulnerabilities for projects they own, easily configurable platform security features are generally underused~\cite{ayala2024secfeat}.
%A general breakdown of the 1,920 projects described in Section~\ref{sec:recruit} is shown in~\ref{tab:underuse}.

\vspace{-0.2cm}

\section{Results}\label{sec:results}

In the following section, we report and discuss the results for 22 semi-structured interviews with open-source maintainers.
In our reporting, we adhere to the structure of the interview guide described in Section~\ref{sec:interviewstudy}.
We report quotes as transcribed with minor grammatical corrections and omissions. e.g., using \textit{``[...]''}, for readability.

We identified \codenums factors from OSS project maintainers, whose projects have a history of vulnerabilities, regarding their vulnerability management efforts using five categories: current practices, general challenges, platform security feature challenges, platform security feature barriers, and platform security feature wants. The curated codes can be found in~\autoref{tab:bigasstable} and list relevant prior work for each code, if applicable. 
 
Our study is closely related to a recent paper that investigated OSS security and trust practices of contributors, owners, and maintainers~\cite{wermke2022qual-OSSP}, i.e., general OSS stakeholders, regardless of vulnerability history or contributions; however,
\textbf{we are the first to investigate vulnerability management challenges that OSS maintainers, whose projects have a history of patched vulnerabilities, face regarding platform security features involving the GitHub Advisory Database}, i.e., by recruiting maintainers whose projects are linked from GitHub security advisories.
This particular perspective is important as OSS project maintainers typically do not have a security background, so learning from maintainers who have a history of vulnerability triaging provides a glimpse of progress towards security-awareness trickling throughout the OSS ecosystem.
Further, using recommended platform security features and practices are key components of transparency and vulnerability awareness, e.g., by publishing security advisories after patching vulnerabilities, for OSS stakeholders. 
We hope to learn more about \textit{how} and \textit{why} such adoption is present, or not, in previously vulnerable projects.

%%%We present how many listing study participants mentioned codes from~\autoref{tab:bigasstable} (\textit{\textbf{L}}) and how many interviewees discussed or argued a particular topic in detail (\textit{\textbf{I}}). 

%\input{content/table_nocolors}
%\input{content/table_original}
\begin{table*}[ht!]
\centering
\begin{adjustbox}{width=\textwidth}
\begin{tabular}{p{0.15cm}lllp{2.68cm}rl}
\toprule
 & \multicolumn{1}{l}{\textbf{F\#}} & \multicolumn{1}{l}{\textbf{Factors and tooling}} & \multicolumn{1}{l}{\textbf{Description}} & \multicolumn{1}{l}{\textbf{RW}} & \multicolumn{1}{r}{\textbf{L}} & \multicolumn{1}{l}{\textbf{I}} \\
\midrule
\parbox[t]{1mm}{\multirow{8}{*}{\rotatebox[origin=c]{90}{\textbf{Current practices}}}}
  & F1 & \textbf{Email or mailing list} & Uses a private email or mailing list for vulnerabilities. & \cite{wermke2022qual-OSSP,lee2017understanding}  & 50 & 11 \\
  & \cellcolor{grey1}F2 & \cellcolor{grey1}\textbf{Proactive disclosure} & \cellcolor{grey1}Established disclosure process after a vulnerability patch. & \cellcolor{grey1}\cite{wermke2022qual-OSSP} & \cellcolor{grey1}46 & \cellcolor{grey1}7 \\
  & F3 & \textbf{Security policy} & Instructs users how to report vulnerabilities. & \cite{wermke2022qual-OSSP} & 37 & 10 \\
  & \cellcolor{grey1}F4 & \cellcolor{grey1}\textbf{GitHub private reporting} & \cellcolor{grey1}A built-in feature for privately reporting vulnerabilities. & \cellcolor{grey1} & \cellcolor{grey1}20 & \cellcolor{grey1}9 \\
  & F5 & \textbf{Automation tooling} & E.g., Dependabot, code scanning, secret scanning. & \cite{onsori2024quantifying,dietrich2023security,he2023automating,mohayeji2023investigating,alfadel2021use,mirhosseini2017can} & 16 & 12 \\
  & \cellcolor{grey1}F6 & \cellcolor{grey1}\textbf{GitHub Issues} & \cellcolor{grey1}Vulnerabilities are reported publicly with GitHub Issues. & \cellcolor{grey1}\cite{preceptionofoss} & \cellcolor{grey1}9 & \cellcolor{grey1}6 \\
  & F7 & \textbf{External tooling} & Utilizes tooling or reports outside of GitHub. & \cite{he2023automating,wermke2022qual-OSSP,fischer2023effectiveness}  & 3 & 2 \\
  & \cellcolor{grey1}F8 & \cellcolor{grey1}\textbf{Ignores vulnerabilties} & \cellcolor{grey1}Ignores vulnerability reports or does nothing in response. & \cellcolor{grey1} & \cellcolor{grey1}2 & \cellcolor{grey1}-- \\
\midrule
\parbox[t]{1mm}{\multirow{9}{*}{\rotatebox[origin=c]{90}{\textbf{General challenges}}}}
  & F9 & \textbf{Supply chain trust} & E.g., waiting for upstream dependencies to deploy a fix. & \cite{alomar2020you,wermke2022qual-OSSP,dabbish2012social,okafor2022soksssc,fourne2023s} & 36 & 6 \\
  & \cellcolor{grey1}F10 & \cellcolor{grey1}\textbf{Lack of understanding} & \cellcolor{grey1}E.g., knowledge gaps, how to start developing a patch. & \cellcolor{grey1}\cite{alomar2020you,anon2025deepdive} & \cellcolor{grey1}35 & \cellcolor{grey1}12 \\
  & F11 & \textbf{Lack of time} & Balancing vulnerability priority with other OSS tasks. & \cite{lawsonfeel,wermke2022qual-OSSP,anon2025deepdive} & 25 & 9 \\
  & \cellcolor{grey1}F12 & \cellcolor{grey1}\textbf{Lack of resources} & \cellcolor{grey1}E.g., \textit{``I do almost everything myself''} (P4). & \cellcolor{grey1}\cite{wermke2022qual-OSSP,canfora2020process} & \cellcolor{grey1}23 & \cellcolor{grey1}11 \\
  & F13 & \textbf{CVE relationships} & Negative experiences with CVE-assigned vulnerabilities. & & 8 & 9 \\
  & \cellcolor{grey1}F14 & \cellcolor{grey1}\textbf{Lack of procedure} & \cellcolor{grey1}Maintainers are not sure how to deal with vulnerabilities. & \cellcolor{grey1}\cite{wermke2022qual-OSSP} & \cellcolor{grey1}7 & \cellcolor{grey1}5 \\
  & F15 & \textbf{Disclosure coordination} & How to reach out to all users and what to disclose. & \cite{alomar2020you} & 7 & 4 \\
  & \cellcolor{grey1}F16 & \cellcolor{grey1}\textbf{Negative attitudes} & \cellcolor{grey1}Past negative interactions with reporters or companies. & \cellcolor{grey1}\cite{miller2022did,raman2020burnout,windowshit,nodeleave,lawsonfeel,destefanis2016software} & \cellcolor{grey1}6 & \cellcolor{grey1}9 \\
\midrule
\parbox[t]{1mm}{\multirow{8}{*}{\rotatebox[origin=c]{90}{\textbf{PSF challenges}}}}
  & F17 & \textbf{Not enough automation} & More automation needed in PSF functionality. & \cite{kabadi2023apr,bui2024apr} & 39 & 11 \\
  & \cellcolor{grey1}F18 & \cellcolor{grey1}\textbf{Too much noise} & \cellcolor{grey1}E.g., too many false positives. & \cellcolor{grey1}\cite{wermke2022qual-OSSP,wedyan2009effectiveness,he2023automating,mirhosseini2017can,wessel2021don,wessel2022bots,brown2019sorry,zampetti2017open} & \cellcolor{grey1}24 & \cellcolor{grey1}10 \\
  & F19 & \textbf{Vulnerability scoring} & E.g., having a hard time calculating CVSS scores. & & 21 & 9 \\
  & \cellcolor{grey1}F20 & \cellcolor{grey1}\textbf{CI processes are missing} & \cellcolor{grey1}GitHub Actions \& tests are not available in private forks. & \cellcolor{grey1} & \cellcolor{grey1}8 & \cellcolor{grey1}7 \\
  & F21 & \textbf{Broken tests \& builds} & Failing builds \& tests after vulnerability patch is applied. & \cite{zampetti2017open,rausch2017empirical,mirhosseini2017can,brown2019sorry} & 6 & 7 \\
  & \cellcolor{grey1}F22 & \cellcolor{grey1}\textbf{Cluttered notifications} & \cellcolor{grey1}E.g., vulnerable dependencies presentation is cluttered. & \cellcolor{grey1} & \cellcolor{grey1}4 & \cellcolor{grey1}4 \\
  & F23 & \textbf{Too much manual setup} & E.g., manually enabling PSFs on all projects. & & 3 & 3 \\
\midrule
\parbox[t]{1mm}{\multirow{7}{*}{\rotatebox[origin=c]{90}{\textbf{PSF barriers}}}}
  & \cellcolor{grey1}F24 & \cellcolor{grey1}\textbf{Lack of awareness} & \cellcolor{grey1}E.g., not knowing about private vulnerability reporting. & \cellcolor{grey1}\cite{ayala2024secfeat} & \cellcolor{grey1}33 & \cellcolor{grey1}12 \\
  & F25 & \textbf{Complex to set up or use} & PSFs are too complex for maintainers with limited time. & & 26 & 7 \\
  & \cellcolor{grey1}F26 & \cellcolor{grey1}\textbf{They are unnecessary} & \cellcolor{grey1}Perception that using PSFs is not needed. & \cellcolor{grey1} & \cellcolor{grey1}23 & \cellcolor{grey1}6 \\
  & F27 & \textbf{Reputation concerns} & Past vulnerabilities reflect a negative project reputation. & & 11 & 4 \\
  & \cellcolor{grey1}F28 & \cellcolor{grey1}\textbf{Bad UI presentation} & \cellcolor{grey1}Hard to find or hidden, \textit{``second-class features''} (P10). & \cellcolor{grey1}\cite{danilova2020one,johnson2013don,fischer2023effectiveness} & \cellcolor{grey1}8 & \cellcolor{grey1}6 \\
  & F29 & \textbf{Lack of motivation} & Feeling burned out from other OSS tasks. & \cite{raman2020burnout, openssfwgs, linåker2024sustaining, amit2024motivation,lawsonfeel} & 5 & 4 \\
  & \cellcolor{grey1}F30 & \cellcolor{grey1}\textbf{Not sure what they do} & \cellcolor{grey1}Unsure of PSFs' functionality, benefits, etc. & \cellcolor{grey1} & \cellcolor{grey1}4 & \cellcolor{grey1}2 \\
\midrule
\parbox[t]{1mm}{\multirow{7}{*}{\rotatebox[origin=c]{90}{\textbf{Maintainer wants}}}}
  & F31 & \textbf{Assisted triaging} & Automation for impact analysis, patch development, etc. & \cite{kabadi2023apr,bui2024apr,wessel2022bots} & 39 & 9 \\   
  & \cellcolor{grey1}F32 & \cellcolor{grey1}\textbf{Assisted PSF setup} & \cellcolor{grey1}More guidance for the PSF setup process. & \cellcolor{grey1} & \cellcolor{grey1}37 & \cellcolor{grey1}10 \\
  & F33 & \textbf{Security-specific funding} & Opportunities for funding OSS security efforts. & \cite{wermke2022qual-OSSP} & 32 & 7 \\
  & \cellcolor{grey1}F34 & \cellcolor{grey1}\textbf{User-friendly resources} & \cellcolor{grey1}Tailored for those without a security background. & \cellcolor{grey1} & \cellcolor{grey1}17 & \cellcolor{grey1}8 \\
  & F35 & \textbf{Gamification for projects} & E.g., a green shield for projects with proper PSFs set up. & \cite{moldon2021gamification,trockman2018adding,dabbish2012social} & 13 & 6 \\
  & \cellcolor{grey1}F36 & \cellcolor{grey1}\textbf{PSF checklist} & \cellcolor{grey1}To-do list for enabling easily configurable PSFs. & \cellcolor{grey1} & \cellcolor{grey1}15 & \cellcolor{grey1}3 \\   
  & F37 & \textbf{Nudges or reminders} & Nudges reminding maintainers to consider using PSFs. & & 12 & 5 \\
\bottomrule
\end{tabular}
\end{adjustbox}
\caption{\codenums factors and tooling codes, shortened to preserve space, identified from the listing survey study and interview study divided into five categories as a result of software vulnerability management efforts. Acronyms listed represent, respectively: platform security features (\textbf{PSF}), existing related work (\textbf{RW}), the number of listing (\textbf{L}) survey study mentions ($n_1=80$), and the number of interview (\textbf{I}) study mentions ($n_2=22$). The extended table without omitted codes can be found in our artifact~\cite{icse25repo}.}
\label{tab:bigasstable}
\end{table*}

\vspace{-0.2cm}

\subsection{Current practices used for vulnerability management}\label{sec:currmanagement}

Both of our studies reflect that OSS project maintainers use a variety of tooling both in and out of the GitHub platform.
Most encourage using a private avenue for reporting vulnerabilities while some are okay with using public channels, e.g., GitHub issues, for security bugs. 
Two listing study respondents indicated that they ignore vulnerabilties altogether.

\vspace{0.1cm}

\noindent\textbf{Emails and external tooling or reports % \diffadd{
(F1 \& F7)} %}
Using a form of a security contact email or mailing list consisting of project maintainers was the most listed form of current vulnerability management practices ($L=50$, $I=11$). This is consistent with a finding from Wermke et al., where authors identified that having a security contact point was the commonly mentioned aspect of a security policy~\cite{wermke2022qual-OSSP}.

Participants indicated that using email for security reports is sufficient because it discourages reporters from creating a public issue and acts as a private avenue for report resolution ($L=11$, $I=4$), e.g., \textit{``we'll take time to really have a discussion and not just here's my report, okay, I dismiss it or I accept it. It's more like a conversation''} (P4). Others mention having security experts on their maintenance team ($L=3$).% Three interviewees shared a perspective that embodies trust among maintainers:

\leftskip=0.5cm\rightskip=0.5cm

\noindent \textit{``With the projects I've been a part of, we just use one email address everybody can log into, which I suppose is just not a super secure way of managing that point of contact, but we all trust each other as maintainers. We have like a shared password on that email account, and that's just kind of how we do things.''} (P12).

\leftskip=0cm\rightskip=0cm

Listing participants and interviewees also mentioned using tooling outside of GitHub, including internal tools for static analysis ($L=2, I=3$), e.g., Coverity~\cite{coverity}, and receiving external reports from companies ($L=1, I=2$). One interviewee described their relationship with a company that sends bundled vulnerability reports for their OSS project semi-annually:

\leftskip=0.5cm\rightskip=0.5cm

\noindent \textit{``They send me a lot of bugs I never knew about and they give me time to fix it. Once fixed, they publish, and then they come back next year [...] I'm surprised with what they can find, I would like to know why they do that but I never complain. Be grateful, you know?''} (P17).

\leftskip=0cm\rightskip=0cm 

%\vspace{0.05cm}

\noindent\textbf{Proactive disclosure process % \diffadd{
(F2)} %}
Having an established disclosure process after patching vulnerabilities consisting of project maintainers was the second-highest form of current vulnerability management practices ($L=46$, $I=7$). This demonstrates a sense of responsibility within the OSS community since it ensures affected dependent client projects within reach are notified to upgrade~\cite{mohayeji2023investigating}. %\jessy{there's a dependabot study that can be referenced here}

\vspace{-0.05cm}

Listing participants use different mediums to let the community know about the need to upgrade, which include sending messages to mailing lists and backchannels ($L=7$), creating GitHub security advisories ($L=6$), and in some cases, quickly requesting a CVE ($L=5$). Interviewee participants were also wary of making sure users are aware of vulnerabilities ($I=7$), two of which felt that creating a CVE is enough to ensure visibility to users and dependents. One interviewee felt that disclosure is only necessary when a vulnerability is particularly important because there is \textit{``a lot of bandwidth on security disclosures to begin with.''} (P6).

\vspace{0.1cm}

\noindent\textbf{Security policies % \diffadd{
(F3)} %}
The next current practice most listed is having a project security policy ($L=37, I=10$). OSS security policies have mostly been used to inform reporters how to properly communicate vulnerabilities to maintainers via security contact~\cite{wermke2022qual-OSSP}, but they are generally underused~\cite{ayala2023workflowssps}. 

Although most listing participants mention using security policies to simply provide a security contact ($L=11$), in some cases, maintainers link an organization-specific security policy published outside of GitHub ($L=3$). One interviewee mentioned how they use security policies to explicitly discourage contributors from reporting vulnerabilities publicly and provide multiple methods of private communication:

\leftskip=0.5cm\rightskip=0.5cm

\noindent\textit{``We have a security policy in place where we say please do not report it publicly but try to contact me personally via email or send a mail to our security mailing list or create a security advisory on GitHub.''} (P13).

\leftskip=0cm\rightskip=0cm

\vspace{0.1cm}

\noindent\textbf{Private vulnerability reporting % \diffadd{
(F4)} %}
Another GitHub platform security feature listing participants mentioned frequently is private vulnerability reporting ($L=20, I=9$). This particular feature allows contributors to privately report vulnerabilities within the GitHub platform. OSS maintainers can then review a report with functionalities to update its severity, invite others to develop a fix, and decide whether or not to request a CVE. 

Listing survey participants and interviewees mention using the private forks from private vulnerability reporting to quietly develop fixes ($L=6, I=7$). Four interviewees mentioned how they like this feature because it provides an easy method of reporting security bugs and is quick to set up, i.e., by clicking enable. One interviewee describes private vulnerability reporting as \textit{``very comfortable and very easy to use.''} (P9). Others expressed satisfaction with private vulnerability reporting because everything is in a central space ($I=2$), i.e., submitted reports are contained in GitHub.

\vspace{0.1cm}

\noindent\textbf{Automation tooling % \diffadd{
(F5)} %}
Another method of current practices listing and interviewees mention is the usage of automated dependency analysis tooling and code scanning $(L=16, I=12)$. Dependency analysis provides alerts of library upgrades, e.g., patched versions, while code scanning can reveal vulnerabilities, e.g., secrets, as a result of running static analysis.

Listing participants and interviewees mentioned using Dependabot as a primary source of upgrading dependencies ($L=12, I=6$). Others mentioned using the Renovate bot that can allow dependencies to be auto-merged if there are no code conflicts ($L=4, I=3$), reducing manual overhead:

\leftskip=0.5cm\rightskip=0.5cm

\noindent\textit{``The first one that I use quite often is Renovate, that is a tool in Github easily available where you can configure: I want this and this upgraded like that, and you can have all kinds of settings and then it automatically gives you a pull request [...] with a dependency update and automatically, the test pipeline fires.''} (P1).

\leftskip=0cm\rightskip=0cm

However, participants describe how platform security features, which are recommended for effective vulnerability management, pose still too much manual effort (Section~\ref{sec:psfchallenges}).

\vspace{0.05cm}

\noindent\textbf{GitHub Issues % \diffadd{
(F6)} %}
Some listing participants and interviewees indicate using GitHub issues for reporting vulnerabilities ($L=9, I=6$). When a GitHub Issue is submitted, it is publicly accessible in the Issues tab located at the top of the project landing page. 

Two listing participants in particular do not mind having vulnerabilities being reported publicly because of visibility, while one listing participant described how they use \textit{``the same public PR process I use for any other issue.''} One interviewee described how they do not mind using GitHub issues when a security bug is not determined to be severe to get help from the public, while another interviewee explained their rationale: 

\leftskip=0.5cm\rightskip=0.5cm

\noindent\textit{``If somebody reports a security issue [publicly], I don't really see a problem with it. Obviously it can be problematic, but I don't really care that much about how it is reported as long as I fix it in a short time.''} (P19).

\leftskip=0cm\rightskip=0cm

\vspace{0.05cm}

%These perspectives reflect that while public issue reporting can enhance transparency and community support, private methods might still be preferred for handling more severe vulnerabilities, especially for projects gaining popularity.

\noindent\textbf{Ignoring vulnerabilties % \diffadd{
(F8)} %}
Two listing participants mention how they ignore vulnerabilities. One attributed this behavior to a lack of motivation for setting up any security features or tooling and indicated that \textit{``financial support or sponsorship''} would be most beneficial for them to improve their vulnerability management practices. The other respondent listed \textit{``unless the project is extremely sensitive - avoid implementing any policy''} and expressed the perception that having security features enabled for projects is unnecessary. Though this respondent listed that they were interested in participating in our interview study, they did not respond with any contact information or reach out to us directly. 

\vspace{-0.1cm}

\begin{finding}[label=find:sec_adv_review_rate]\label{t1}
%\diffadd{
%Most OSS maintainers in our study use designated emails, platform security features, or automation tooling for private vulnerability reporting and discovery. \st{Others mention choosing to deal with reported vulnerabilities publicly, or ignoring such vulnerabilities altogether.} \josh{I'm unsure why we are including this second sentence as a takeaway. What insight is gained by saying to anyone that a vulnerability can either be dealt with or ignored?}
%Those who publicly handle or ignore vulnerabilities can expose projects and users to security risks, e.g., early exploits.
Most OSS maintainers in our study take vulnerability management seriously, citing dedicated emails for private vulnerability reporting and established disclosure processes for informing affected projects after patching. However, built-in PSFs are not primarily used, indicating a need for usability improvement and further awareness.
%}
\end{finding}

\vspace{-0.4cm}

\subsection{Challenges with vulnerability management}\label{sec:genchallenges}

All interviewees described wearing many hats when it comes to OSS maintenance, some of who are lone maintainers ($I=11$).
General vulnerability management challenges OSS maintainers face range from trusting the software supply chain to lack of time and resources to issues with CVEs.
%Many participants noted how bug bounty reports are just security-oriented pull requests. 
Participants also describe how a lack of standardized vulnerability handling procedures and implementing proper coordinated disclosure are prominent challenges.
Some bring up the idea of having \textit{``imposter syndrome''} when receiving vulnerability reports, while others share negative reporter experiences. 

\vspace{0.1cm}

\noindent\textbf{Supply chain trustworthiness % \diffadd{
(F9)} %}
Supply chain trust was the most listed challenge with vulnerability management ($L=36, I=6$). Use cases and duties related to the software supply chain in OSS include adopting upstream dependencies, tracking the status of dependencies, and updating dependencies in a timely manner, e.g., when a vulnerability is patched.

Specific challenges that the participants mentioned include the burden of keeping updated with dependencies and the latest vulnerabilities ($L=20$) and dealing with unmaintained dependencies or delays in pushing a vulnerability fix ($L=12$), which can be time-consuming and resource-intensive. 
One interviewee was particularly concerned about scenarios \textit{``where people purposefully like to put security vulnerabilities into [OSS projects]''} (P6), highlighting the risk of malicious actors deliberately introducing vulnerabilities into upstream dependencies, e.g., the \code{xz-utils} incident (CVE-2024-3094)~\cite{xzutils}.

\vspace{0.1cm}

\noindent\textbf{Lack of understanding % \diffadd{
(F10)} %}
The second most listed challenge for conducting vulnerability management is a lack of understanding ($L=35, I=12$). 
Not being able to effectively understand reported vulnerabilities can lead to delays in patching and disclosure to affected dependent client projects.

Factors reported affecting a lack of understanding include complexity ($L=19, I=4$), developing a patch ($L=7, I=7$), testing ($L=6, I=6$), and knowledge gaps ($L=5, I=8$). 
One participant listed that \textit{``some vulnerabilities can be complex and require extensive investigation and testing to ensure they are fully resolved without introducing new issues,''} reflecting platform security feature challenges we explore in Section~\ref{sec:psfchallenges}, while one interviewee said \textit{``my brain is far too small to understand [vulnerabilities]''} (P9). We explore potential avenues maintainers believe OSS platforms can take to minimize such knowledge gaps and awareness in Section~\ref{sec:psfwants}.

\vspace{0.1cm}

\noindent\textbf{Lack of time and resources % \diffadd{
(F11 \& F12)} %}
The next most listed software vulnerability management challenges are the lack of time ($L=25, I=9$) and resources ($L=23, I=11$). The time OSS maintainers spend on developing and managing projects is limited, e.g., hobbyists. Further, resources are scarce, e.g., large-scale software ecosystems like Python Package Index and NPM are primarily made up of projects with a single maintainer~\cite{npmgraph,pypigraph}.

Challenges associated with a lack of time and resources include balancing prioritizing vulnerabilities with ongoing development tasks ($L=21, I=16$), the ability to triage reports promptly ($L=15, I=11$), and getting help from others ($L=8, I=5$). One interview participant summarizes:

\leftskip=0.5cm\rightskip=0.5cm

\noindent\textit{``While I prioritize addressing vulnerabilities immediately, balancing this with ongoing development tasks can sometimes be challenging [...] There are limited resources and time available to address every reported vulnerability quickly.''} (P17).

\leftskip=0cm\rightskip=0cm

\vspace{0.1cm}

\noindent\textbf{Negative relationships with CVEs % \diffadd{
(F13)} %}
Participants also expressed bad relationships with CVEs and the overall CVE process as a challenge when conducting vulnerability management ($L=8, I=9$). 
CVEs play an important role in the overall disclosure process and help promote transparency in the OSS ecosystem, e.g., upgrade notifications. 

One listing participant explains their stance on CVEs: 

\leftskip=0.5cm\rightskip=0.5cm
\noindent\textit{``The CVE program suffers from many single points of failures: managed by the USA (not 24/7) hence a CVE ID cannot be delivered fast. CISA analysts backlog and [don't] have enough time and understanding of the system's complexity to properly analyze reports; thus, publish poor quality content. The whole process is outdated and must be reformed to a distributed approach to enable international sources of trust to work.''}

\leftskip=0cm\rightskip=0cm

Others described experiences where fixes were put on hold because of pending CVEs ($L=3$) or feel intimidated by the nature of CVEs ($L=2,I=2$). In particular, one interviewee described their negative perception of CVEs and observation of other project maintainers' handling of CVEs:

\leftskip=0.5cm\rightskip=0.5cm

\noindent\textit{``I'm incentivized to lie to make the CVE [severity] lower because it makes my project look bad, you have to be really, really honest [...] I noticed a lot of people like downgrade their CVEs.''} (P15).

\leftskip=0cm\rightskip=0cm

This quote highlights potential ethical dilemmas and pressures faced by OSS maintainers when dealing with CVEs, which can negatively affect the accuracy and reliability of severity metrics throughout the OSS ecosystem.

\noindent\textbf{Lack of procedures and coordinated disclosure % \diffadd{
(F14 \& F15)} %}
Though established procedures can help alleviate the receive-to-resolve timeline, the lack of procedures is a challenge OSS maintainers also face ($L=7, I=5$). Further, participants also express challenges with being able to effectively reach affected dependent projects and users ($L=7, I=4$).

OSS maintainers express a lack of recommended standardized processes for handling vulnerabilities ($L=5, I=1$) due to limited experience ($L=2$), some even feeling intimidated ($L=2$). 
Two interviewees described a sense of imposter syndrome when receiving vulnerabilities because \textit{``there are so many things we need to watch out for, it's hard to stay confident that you're doing the right thing.''} (P13). 
Further, participants expressed concerns about effectively disclosing the nature of the vulnerabilities ($L=4, I=7$) and a sense of mistrust with upstream projects not disclosing vulnerabilities properly ($L=2, I=2$). 
These perspectives suggest needs for clearer guidelines and better support for OSS maintainers in handling vulnerabilities, reported in Section~\ref{sec:discussion}, as well as further research on the adoption of and improving attestation practices, e.g., software-bill-of-materials, to foster upstream trust within the software supply chain~\cite{donoghue2024sbom,xia2023sbom,nocera2023sbom}.

\vspace{0.1cm}

\noindent\textbf{Negative attitudes (F16)} % \diffadd{(F16)} %}
The least listed challenge of vulnerability management is negative attitudes from reporters ($L=6,I=9$). The OSS ecosystem is human-centered, e.g., via community and collaboration, and is not new to toxic interactions~\cite{raman2020burnout,miller2022did}. One interviewee described experiences with companies demanding pentests. Participants mention the argumentative nature with vulnerability reporters ($L=3, I=7$) and those with sole intentions to make money ($L=3, I=5$).

\leftskip=0.5cm\rightskip=0.5cm

\noindent\textit{``They give me this list of vulnerabilities in an email and then say, we're going to make it public in a week. And I think it's not so bad to make it public; in fact, both times I said go ahead and make it public right now. But the way people give you this problem and say fix it or else, it's not a very conducive environment.''} (P2).

\leftskip=0cm\rightskip=0cm

\vspace{-0.1cm}

\begin{finding}\label{t2}
%\diffadd{
%Aside from knowledge gaps and time/resource constraints, %as particularly challenging when handling vulnerabilities, which hinder their ability to effectively address security issues. 
%\st{Others describe negative experiences with CVEs, e.g., ethical dilemmas, and reporters/companies, e.g., demanding fixes.} \josh{Why is there an ``, e.g.," here? It reads weird.}
Aside from knowledge gaps and lack of time/resources, most OSS maintainers in our study are wary of adopting dependencies due to supply chain risks.
Ethical dilemmas around disclosure and reporters/companies demanding fixes further complicate OSS responsibilities.
%} %pressure from 
\end{finding}

\vspace{-0.4cm}

\subsection{Challenges with platform security features}\label{sec:psfchallenges}

The adoption of platform security features (PSFs) promotes a robust vulnerability management process, especially if they are usable for OSS maintainers without a security background. 
In this section, we report various challenges and first-hand experiences as a result of adopting such features, revealing why some maintainers may choose to disable them as a result. 

\vspace{0.1cm}

\noindent\textbf{Not enough automation (F17)} % \diffadd{(F17)} %}
The most listed PSF challenge described is that there is not enough automation ($L=39, I=11$). 
As described earlier in Section~\ref{sec:genchallenges}, OSS maintainers have limited time and resources to address every reported vulnerability quickly.
Automation for security-oriented tasks helps promote easy security without additional overhead.

Participants express challenges regarding automation to update dependencies ($L=24, I=7$). In particular, five interviewees are okay with auto-merging dependency upgrades as much as possible, while others would prefer to do so on a library-by-library basis ($I=2$). One interview participant mentioned auto-merging private vulnerability report fixes if the build does not break; however, private forks do not have CI integration and participants describe experiences with broken builds and tests, which we report later in this section.

One interviewee describes how they are okay with merging automated pull requests regardless of testing: \textit{``Go ahead and merge the PR. I don't even really test it.''} (P5). This perspective reflects a desire for streamlined processes, even if it comes at the cost of rigorous testing, underscoring the tension between automation convenience and careful oversight needs.

%\vspace{0.1cm}

\noindent\textbf{Too much noise and vulnerability scoring issues % \diffadd{
(F18 \& F19)} %}
Listing participants mentioned that too much noise from PSFs ($L=24, I=10$) and issues from vulnerability scores ($L=21, I=9$), e.g., inaccuracy, are particularly challenging to deal with. Participants describe noise as many false alarms, different from spam or low-quality reports, and how noise can take time away from general OSS tasks and in some cases, overwhelm or scare OSS maintainers ($L=4, I=6$). 

It is common knowledge that noises from dependency management~\cite{mirhosseini2017can,he2023automating} and static analysis~\cite{wedyan2009effectiveness,johnson2013don} tools are problematic.
A majority of noise mentioned is from dependency false positives ($L=8, I=6$), while others said it is from static analysis tooling ($L=4, I=2$), e.g., code scanning, and a general sense of annoyance from notifications ($L=2, I=2$). One interviewee mentioned how even proprietary scanners produce too much noise -- we report remediations to help improve PSF setup and reduce noise in Section~\ref{sec:psfwants}. Two interviewees mentioned how noise negatively impacts their motivation to continue maintaining OSS projects, one described:

\leftskip=0.5cm\rightskip=0.5cm

\noindent\textit{``I left this project, I'm not doing this anymore [...] I think there's a lot of noise [...] and then the dedication, the love and the passion, the patience, going over it, and taking care of it, it's not easy at all. So security is kind of a second thought to most of us.''} (P8).

\leftskip=0cm\rightskip=0cm

In addition to noise, participants describe issues with vulnerability scores from reported vulnerabilities, e.g., CVE scores are inflated ($L=5, I=4)$ and unsure how to correctly calculate or adjust CVSS scores ($L=3, I=8$). One interview participant summarizes their past experiences with attempting to properly calculate security metrics:

\leftskip=0.5cm\rightskip=0.5cm

\noindent\textit{``I always have issues with calculating CVSS [scores]. All the documentation I found seems to be like for security researchers, or people who have experience with security incidents. There's nothing that just describes it in everyday terms that I can really wrap my head around, so that's one of the biggest difficulties.''} (P3).

\leftskip=0cm\rightskip=0cm

%\vspace{0.1cm}

\noindent\textbf{Absent CI processes in private forks, failing tests and broken builds in the public eye % \diffadd{
(F20 \& F21)} %}
The next most listed challenge with PSFs is the lack of CI processes when developing fixes on a private fork ($L=8, I=7$), i.e., as a result of adopting the built-in private vulnerability reporting feature. Although private forks are recommended for working on fixes with reporters, not being able to run CI processes on such fixes in the same environment as merging pull requests on the main branch can be problematic, resulting in overhead.

To get around the lack of CI processes in private forks, participants have resorted to hosting additional private repositories outside of GitHub that mimic their public presence ($L=2, I=1$) or running potential fixes through build processes on personal machines ($I=4$). 
Both of these solutions, while functional, introduce additional challenges and complexities, e.g., maintaining parallel environments and ensuring consistency across different setups.
One interview participant described an elaborate approach: 

\leftskip=0.5cm\rightskip=0.5cm

\noindent\textit{``We have [someone], our traffic controller, and their role is to check that the incoming security reports [...] If they consider it something we need to be concerned with, they create an issue in a private project [...] It really slows the process down, because we have to merge one patch, then we have to go to our repo, pull in commits to the new one we need to release.''} (P14).

\leftskip=0cm\rightskip=0cm

Participants report experiences of failed tests or broken builds after patches are merged to the main public repository ($L=6, I=7$), so OSS maintainers must thoroughly ensure fixes do not introduce new issues. 
\textit{``The biggest problem for us is the fact we can't run our automated tests on the fixes that we make on those forks.''} (P14).
This emphasizes a development workflow issue caused by the lack of automated testing capabilities in private forks, which is crucial for program repair and preventing regressions -- a current multi-dimensional and challenging research area~\cite{kabadi2023apr,bui2024apr,minhas2023regression}.%fu2022vulrep,minhas2023regression,brahneborg2017regression}.

\vspace{0.1cm}

\noindent\textbf{Cluttered notifications, setup and usage is too manual % \diffadd{
(F22 \& F23)} %}
The least listed PSF challenges are cluttered notifications ($L=4, I=4$) and too much manual effort ($L=3, I=3$), i.e., usability challenges.
Interviewees describe that the best security is \textit{``easy security''} ($I=6$), including the ease of use for PSFs. When security features are user-friendly and require minimal effort to manage, they can be effectively utilized by OSS maintainers without additional overhead.

Participants describe how dependency notifications are too cluttered ($L=2, I=3$), and how there should be a merge-all button, i.e., to reduce manual effort, for dependency upgrades that do not cause merge code conflicts ($L=1, I=3$). %One interviewee describes such clutter:

\leftskip=0.5cm\rightskip=0.5cm

\noindent\textit{``They are all grouped together at the bottom of your notification panel and they appear only if all the others are marked as done [...] just terrible to deal with [...] There is like a cron job that runs then delivers you all of those once a week, they're all bundled.''} (P10).

\leftskip=0cm\rightskip=0cm

Other tasks described as too manual by interviewees include having to individually enable PSFs for each project ($I=4$), e.g., as opposed to a centralized approach where all projects are presented on a single screen, and having to manually add the same collaborators to a private fork every time a vulnerability is privately reported ($I=3$). 

\vspace{-0.1cm}

\begin{finding}\label{t3}
%\diffadd{ %\st{Most OSS maintainers describe that PSFs lack automation, cause too much noise, and do not provide adequate help for vulnerability scoring.} \josh{It looks like you are just adding takeaways that repeat the bolded subheaders of this section. That does not come off as insightful to me. I'm concerned the most negative reviewer will likely agree.} 
PSFs often rely on manual intervention and are noisy, requiring maintainers to sift through and triage security risks without sufficient automated support. Further, some PSFs are missing core CI processes, cause broken tests/builds, and are not UI-friendly, resulting in overhead.%\josh{This second sentence is insightful, but the previous sentence is not.}
%}
\end{finding}

\vspace{-0.4cm}

\subsection{Barriers hindering adoption of platform security features}\label{sec:psfbarriers}

Despite the importance of actively practicing secure coding practices and being knowledgeable about general security concerns, OSS developers typically do not have a security background.
A majority of our participants also lack such background ($L=50,I=15$). 
In this section, we report possible barriers that hinder PSF adoption, including lack of awareness, complex setup procedures, concerns about project reputation, and perceptions that they are unnecessary.\\

\vspace{-0.3cm}

\noindent\textbf{Lack of awareness and bad UI presentation % \diffadd{
(F24 \& F28)} %}
The most listed PSF barrier is lack of awareness ($L=33, I=12$). PSFs are available for usage so that OSS projects can have robust vulnerability management processes and proper disclosure upon vulnerability patching. This lack of awareness implies that although tools and processes are available to help promote a thriving secure OSS ecosystem, they are underutilized as indicated in prior work~\cite{ayala2023workflowssps}.% measuring OSS PSFs' usage~\cite{ayala2023workflowssps}.% and in our recruitment pool statistics from~\autoref{tab:svmusage}.

Participants mentioned how they were not aware of security policies ($L=43, I=8$), private vulnerability reporting ($L=43, I=10$), and public security advisories ($L=34, I=4$) or the GitHub advisory database ($L=13, I=6$).
Further, 26.7\% (21/80) of listing participants indicated that they have none of the three previously mentioned PSFs enabled.  
One interviewee expresses how \textit{``the main problem is just that many people don't know about it''} (P18), and another interviewee, who has had eight CVE-assigned vulnerabilities in a project they manage, expressed curiosity after learning about what private vulnerability reporting has to offer:

\leftskip=0.5cm\rightskip=0.5cm

\noindent\textit{``It's on my list to actually research now. Yeah, it looks like they've got a pathway to disclosure in CVEs that's integrated and seems like a good tool for us.''} (P20).

\leftskip=0cm\rightskip=0cm

Related to lack of awareness, bad UI presentation is another PSF barrier participants mentioned ($L=8, I=6$). In particular, participants feel that PSFs are too hidden or hard to find ($L=5, I=6$) and that there are just too many features to potentially configure ($L=2, I=3$). One interviewee described how PSFs feel like \textit{``second-class features''} (P10). This perception can discourage maintainers from exploring PSFs' capabilities, further contributing to their underuse.

\leftskip=0.5cm\rightskip=0.5cm

\noindent\textit{``I mean, I'm just looking at how to or for a way to roll stuff out, but it's pretty hidden. I can roll out a policy that allows private vulnerability and a lot of other stuff, but you have to really click through and find it.''} (P11).

\leftskip=0cm\rightskip=0cm

%\vspace{0.1cm}

\noindent\textbf{Complex to set up or use, not sure what they do % \diffadd{
(F25 \& F30)} %}
Listing participants mentioned that PSFs are too complex to set up or use ($L=26, I=7$). Though not nearly listed as much, participants are \textit{``not sure what they do in the first place''} ($L=4, I=2$). These perceptions reflect a barrier of adoption since OSS maintainers question their overhead and purpose.
Interviewees express a sense of excessive overhead with setting up PSFs ($I=4$). 
Others cite general ignorance as to why they are not aware of what particular PSFs do ($L=3, I=7$). 
As a result, they did not bother using PSFs.%looking into PSFs.
If users are unclear about what these features do or how they can benefit their projects, they are less likely to invest effort in setting them up, especially if they are perceived as complex.

\leftskip=0.5cm\rightskip=0.5cm

\noindent\textit{``The biggest reason I never used them is they've never been pushed or the benefits of them sold to me [...] If it's really easy and simple to use, it'd be nice if that is kind of turned on by default on all projects.''} (P12).

\leftskip=0cm\rightskip=0cm

Participants also reference previous experiences with trying out PSFs but refuse to look into additional offered tooling because of usage complexity ($L=11, I=4$), e.g., with dependency graphs ($L=2$). These insights suggest simplifying the setup process and providing clearer information about the purpose and use of PSFs for OSS maintainers looking to adopt them to manage vulnerabilities.\\

\vspace{-0.35cm}

\noindent\textbf{Perception that they are unnecessary % \diffadd{
(F26)} %}
The next most listed PSF adoption barrier is the perception that they are unnecessary ($L=23, I=6$). When OSS maintainers perceive PSFs as not adding value to their current processes, they will likely refrain from investing additional time and resources into learning about or implementing these tools; thus, undermining recommended efforts to promote a secure OSS ecosystem.

Participants mention how PSFs are unnecessary for a variety of reasons, including the perception that projects lack importance ($L=16, I=6$), adding PSFs is overkill ($L=5, I=3$), or that maintainers do not value security ($I=2$). These perceptions highlight a need for more documentation on the value of security measures ($L=17, I=8$) and mechanisms for PSFs to be tailored to fit specific project needs and scale.

\leftskip=0.5cm\rightskip=0.5cm

\noindent\textit{``I haven't really needed anything more involved than GitHub issues [...] Security isn't something that we worry too much about. We're not ready to hear that message, even if GitHub does push me, I'll probably just skim over them, because I'm not ready to actually to, you know, get that message [...] We worry, we kind of have it in mind, but it's not our main goal.''} (P8).

\leftskip=0cm\rightskip=0cm

\noindent\textbf{Project reputation and a lack of motivation % \diffadd{
(F27 \& F29)} %}
The remaining barriers to PSF adoption OSS maintainers mention are concerns with project reputation ($L=11, I=4$) and a lack of motivation ($L=5, I=4$). Concerns about maintaining a positive public perception can deter OSS maintainers from implementing PSFs, that would otherwise enhance security, while those without motivation to adopt PSFs may deprioritize or ignore the use of PSFs.
The attitude concerning reputation is not unique to maintainers. Alomar et al.~\cite{alomar2020you} observed this attitude at the organizational level, where organizations try to lower the severity ratings of vulnerabilities in their products.

Participants point out fear of negative project reputation as a reason to not adopt PSFs ($L=3, I=3$), e.g., owning a project with previous high or critical CVE-assigned vulnerabilities ($I=1$). On the other hand, some interviewees consider a project without vulnerabilities as suspicious ($I=2$), i.e., patched vulnerabilities are a good sign of a ``healthy project'' (P14). One interviewee reflected on such a dilemma:

\leftskip=0.5cm\rightskip=0.5cm

\noindent\textit{``It's always in the back of my mind when looking at an issue and seeing, should this go through the security advisory process? Or should it just be a normal PR and fix? That's the end of it. But that's wrong, and I know it. It still feels like, you know, hurting the reputation of my project. But it's wrong, I know it.''} (P3).

\leftskip=0cm\rightskip=0cm

When it comes to lack of motivation, participants cite ``maintainer burnout'' as the primary reason for not wanting to deal with vulnerabilities ($L=3, I=4$). This burnout can lead to a reduced willingness to engage with vulnerabilities, as maintainers may already be overwhelmed by other OSS tasks, e.g., high-stress levels from frequent demands for features and bug fixes ~\cite{raman2020burnout}. Further, others see no benefit for improving project reputation through a security lens ($L=1, I=2$), leading them to deprioritize or disregard the adoption of PSFs.

\vspace{-0.1cm}

\begin{finding}\label{t4}
%\diffadd{%Most OSS maintainers cite awareness issues, setup complexity, and lacking necessity as primary PSF adoption barriers.
Most OSS maintainers are not aware of PSFs; there are too many to look through and understand their relevance.
Further, project reputation concerns, e.g., past vulnerabilities make projects look bad, 
%\josh{Just look bad or make the project look bad?}
and lack of motivation to manage vulnerabilities also hinder PSF adoption.%}
\end{finding}

\vspace{-0.4cm}

\subsection{Opportunities for improvement \& support}\label{sec:psfwants}

Lastly, we asked participants about success stories or positive outcomes as a result of their continuous vulnerability management efforts, if applicable, as well as supplemental features or improvements to current PSFs that would benefit their current vulnerability management approaches the most.

\vspace{0.1cm}

\noindent\textbf{Assisted vulnerability analysis and triaging % \diffadd{
(F31)} %}
Participants expressed the most need for assisted vulnerability analysis and triaging ($L=39, I=9$), i.e., via automation mechanisms. 
In particular, participants felt that this would be useful for understanding vulnerabilities and their actual impact on OSS projects ($L=34, I=7$) to reduce false positives and noise described in Section~\ref{sec:psfchallenges}, e.g., \textit{``I hope a tool can take a context into consideration and it should help tooling avoid the false positives, and to trigger only this stuff when [vulnerable]''} (P7).
Further, participants express a need for assistance with developing fixes and generating tests ($L=21, I=6$), which could streamline the remediation process and ensure more effective vulnerability management, e.g., \textit{``generate, you know, tests with test cases or something like that, maybe as a way to lighten the burden of actually developing tests''} (P5). 

\vspace{0.1cm}

\noindent\textbf{PSF setup assistance, checklists, and nudges % \diffadd{
(F32, F36, \& F37)} %}
Participants also showed interest in having assisted PSF setup ($L=37, I=10$). 
In particular, interviewees mention needs for assistance generating security policy content using project scope ($I=3$), security tooling recommendations based on project contents ($I=6$), and interactive guidance throughout the setup process ($I=7$), e.g., \textit{``it's a little bit not easy to use [...] I need to figure out, okay, what is the proper setup for my project [since] I'm of the mindset that brevity is king.''} (P9). 
Further, participants mention wanting checklists and suggestions of things to do for recommended PSFs ($L=15, I=3$), alongside nudges or reminders and ``easy'' configurations as methods of encouraging others to strengthen their projects' security posture ($L=16, I=5$). One interview elaborates on their view of overall PSFs: \textit{``It's pretty overwhelming, and you might just be like, why do I need any of these? Like, why is this important? Having a single button that just says, enable best practices, would go a long way.''} (P16). 

\vspace{0.1cm}

\noindent\textbf{Security funding and cyber defense gamification % \diffadd{
(F33 \& F35)} %}
The next most-listed want for participants is security-specific funding ($L=32, I=7$), i.e., funding to be used for security-related tasks and efforts. 
In particular, participants mention using funds for maintainers and reporters to get paid ($L=18, I=6$), funding a dedicated bounty pool for valid vulnerabilities ($I=3$), or requesting annual reviews from security experts ($I=2$). 
%%%\textit{``I'm a bit concerned that if I open a door on GitHub for people to report vulnerabilities, I'm a little bit scared that they will ask something in return and I will not be able to compensate them.''} (P22).
There was also interest in cyber defense gamification ($L=13, I=6$), i.e., to recognize projects that adopt vulnerability management processes. 
Four interviewees mention how they use reporter reputation to determine vulnerability report legitimacy. 
Such participants and others expressed interest in project reputation icons, e.g., a \textit{``green shield''} ($I=3$), for projects that adopt recommended PSFs and tooling  ($I=7$); thus, recognizing and rewarding projects that exhibit a proper security posture. %%%One interviewee describes their perspective: 
\textit{``Everyone wants to have the green shield [...] This community health status people see, stuff like that, like you'll be surprised by how many people want to feel those bars.''} (P10). 
%%%Trockman et al.~\cite{trockman2018adding} previously speculated on using gamification for bug fixing.

\vspace{0.1cm}

\noindent\textbf{User-friendly resources and documentation % \diffadd{
(F34)} %}
Lastly, participants expressed interest in having more user-friendly documentation and resources ($L=17, I=8$), some citing general \textit{``ignorance''} of proper OSS security ($I=6$). 
In particular, participants mention OSS recommended training material for \textit{``normie developers''} ($L=3, I=4$), e.g., free webinars ($I=2$), redirection to related projects with implemented security practices ($L=7, I=3$), and \textit{``easier directions''} for setting up and what to expect from PSFs ($I=6$), e.g., not being \textit{``scared''} of false positives ($I=3$). 
\textit{``Security best practices and tooling and all of that is tribal knowledge, and it shouldn't be.''} (P12).

\vspace{-0.1cm}

\begin{finding}\label{t5}
%\diffadd{
Most OSS maintainers want automated vulnerability triaging and PSF support for less overhead when managing vulnerabilities. 
Further, gamification and context-aware checklists can help simplify security efforts and encourage proactive engagement with security best practices.%}
%\josh{Again, just repeating the bolded headers of the section is a waste of space, I think. It's not really addressing what the reviewers are concerned about, as far as I can tell.}
\end{finding}
\vspace{-0.4cm}

\section{Discussion}\label{sec:discussion}

In this section, we present %\diffadd{
additional analyses of the challenges identified and presented in Section~\ref{sec:results},%} 
implications for OSS platforms and OSS maintainers to alleviate prominent challenges we discovered, explore areas for reducing PSF barriers, and expand the benefits participants expressed as helpful for vulnerability management. Further, we present potential research directions aimed at supporting OSS project maintainers during the vulnerability management lifecycle.

\vspace{-0.3cm}

%\diffadd{

\subsection{Bridging factors to reveal underlying issues}

OSS maintainers described challenges from vulnerability management and their experiences with PSFs. 
%\josh{This sentence reads weird. It would be better to enumerate the various angles more rather than just say specific to PSFs or independent of them. At the least, don't end the sentence with ``and more general". That reads extra weird.}. 
Uncovering underlying issues can help identify areas of need, why particular PSFs may not be adopted, and inform future solutions.

%\vspace{0.05cm}
%\noindent\textbf{Knowledge and awareness gaps }

OSS maintainers described challenges that hinder their ability to take necessary actions from PSF output rooted in overwhelming security duties and missing PSF support that would help vulnerability triaging.
Dependency trust issues (F9), e.g., too many false positives from upstream dependencies (F18) and cluttered update notifications (F22 \& F28), cause fatigue (F29) and a disinclination to continue PSF adoption.
Further, maintainers expressed struggle with developing patches (F10, F14, \& F31) since core CI/CD processes are missing from some PSFs (F20) that cause broken tests/builds (F21); thus, leading maintainers to deem PSFs as unnecessary (F26), deploy custom behind-the-scenes processes that can cause burnout (F29), and become overwhelmed in the CVE process (F13, F15, \& F16).
Without any clear prioritization or impact analysis, some PSFs may seem more like distractions, and the lack of trust demonstrated by maintainers is a reinforcing factor in the perception that PSFs are unnecessary, especially if they seem to add little value to the project.
Addressing this concern means incorporating features that would be perceived to provide real value, e.g., timely patching, and the use of gamification or funding (F33 \& F35) and checklists (F36) can also be used to reframe these tasks as rewarding and meaningful, promoting a positive narrative for security.

OSS maintainers also described challenges that hinder their ability to adopt PSFs rooted in misunderstanding and misinterpretation of PSFs' configurations, use cases, and purposes.
Knowledge gaps (F10), e.g., around vulnerability scoring procedures (F19), hinder effective prioritization and remediation efforts. 
Resource constraints (F12), including limited time (F11) and lack of automation (F17), exacerbate these difficulties, often leading to delays in addressing vulnerabilities or adopting PSFs; further, the complexity (F25) and lack of awareness (F24) surrounding PSFs create additional barriers.
If OSS maintainers do not fully understand the benefits or use cases for PSFs (F30), or find them confusing, they are unlikely to explore or enable them (F37). Training resources that simplify security concepts and integrate clear documentation (F32 \& F34) for PSFs could bridge this gap, empowering maintainers to overcome these hurdles and integrate PSFs effectively into their workflow.

\vspace{-0.3cm}

\subsection{Implications \& future directions based on collective challenges}

OSS maintainers mentioned supply chain trust and lack of understanding as the most generally challenging in our listing study, while not enough automation and too much noise were the most listed PSF-specific challenges. 
%\jerry{There are quite a few work on noises because of PSFs (mentioned in related work). Should they be mentioned here too?}

When maintainers wait for upstream dependencies to fix a vulnerability, there is a delay in addressing security issues within their projects, thus exposing OSS projects to potential exploits.
Tools that automate identifying and prioritizing vulnerabilities can help maintainers quickly determine which issues require immediate attention and which can be deferred.
%\diffadd{
To support assisted vulnerability analysis and triaging, tooling should be explored that implements mechanisms for automatically identifying the impact of vulnerabilities using source code as context, i.e., to reduce noise from security tooling.
%Tooling should be explored that implements mechanisms for automatically identifying the impact of vulnerabilities using source code as context, i.e., to reduce noise from security tooling.
To that end, future research can leverage LLMs to (1) help OSS maintainers with interpreting reported vulnerabilities, e.g., by leveraging OSS security datasets like those curated by OpenSSF~\cite{openssfwgs}, and (2) help OSS maintainers generate patches for reported vulnerabilities with minimized regression tests by using the contents of disclosed reports.
% Version A, in comparison to Version B, make it clear that uses of LLMs has it own challenge
% Version A
%\diffadd{
Although LLMs are useful for tasks like code summarization and generation ~\cite{zan2023llms,ahmed2023fewshot}, uses of LLMs pose challenges. For instance, LLMs may misinterpret security reports or generate incomplete/inaccurate patches, leading to regressions; further, there may be hesitance to trust AI-generated suggestions because LLMs’ decision-making processes are uninterpretable~\cite{hou2024llm}.%}
% Version B
%\diffadd{Although LLMs are useful for tasks like code summarization and generation ~\cite{zan2023llms,ahmed2023fewshot}, LLMs may misinterpret security reports or generate incomplete/inaccurate patches, leading to regressions; further, maintainers might be reluctant to trust AI-generated suggestions because LLMs’ decision-making processes are uninterpretable~\cite{hou2024llm}.}
OSS platforms should consider providing CI feature capabilities in private forks so OSS maintainers can expedite fixing vulnerabilities and reduce costs of regression tests, which can be very high~\cite{chittimalli2009recomputing,huang2012history}, indicating a need for research focused on regression testing for security. GitHub disallows private forks from using CI features for security purposes~\cite{ghconcern}, but OSS maintainers desire such features in our studies, especially for fixing vulnerabilities, suggesting the need for further research on secure CI features for vulnerability management.
%\diffadd{
This might involve sandboxing or access controls, e.g., where private forks can use CI features under controlled conditions.%}

Other prominent vulnerability management challenges OSS project maintainers face are negative CVE relationships and vulnerability scoring. 
Collectively, these aspects may lead to undermining or misreporting critical vulnerabilities; as a consequence, polluting the software supply chain with inconsistencies. 
%There is a need to explore solutions that standardize the OSS vulnerability management lifecycle, e.g., developing and evaluating new automated tools and frameworks that provide consistent methods for managing OSS vulnerabilities 
%\josh{This sentence does not seem to follow from the previous sentence, connect well with the following sentence, or draw much from the paper's results, especially if you are arguing for standardization}.
Tooling that can take context into account when a vulnerability exists, including the deployment environment, the project’s purpose, and how the vulnerability interacts with other components, would be especially beneficial for working toward problems in automated vulnerability scoring, which is particularly understudied, as far as we are aware.
%\diffadd{OSS platforms should also consider hosting periodical online webinars, e.g., for topics like vulnerability scoring, tailored for OSS maintainers who are \textit{``normie developers.''}}

%Most interviewees expressed positive attitudes toward integrating large language models during the vulnerability management lifecycle ($I=16$). We encourage researchers to investigate further how LLMs can be leveraged to (1) help OSS maintainers with interpreting reported vulnerabilities, e.g., by leveraging datasets curated by OSS security working groups like OpenSSF~\cite{openssfwgs}, and (2) help OSS maintainers generate patches for reported vulnerabilities with minimized regression tests by using the contents of disclosed reports, e.g., advisories~\cite{dunlap2024vfcfinder}. We also encourage researchers and OSS platforms to develop practical educational resources accessible to developers without a security background, e.g., standardized security guidelines~\cite{openssfblog}, that support OSS vulnerability management efforts to help minimize knowledge gaps.

\vspace{-0.3cm}

\subsection{Implications \& future directions based on PSF adoption barriers}

The most listed PSF barriers by OSS maintainers were lack of awareness, complexity, and perception of being unnecessary were the most prominent barriers; thereby, preventing PSF adoption as a part of vulnerability management processes. 
Notably, of the twelve interview participants who described a lack of PSF awareness, five said they would enable them as a result of just learning about their existence.
%OSS project maintainers should communicate the positive impact of bug bounty programs on OSS so other projects reluctant to participate can be intrigued by the benefits of receiving a report from a third-party program. 

We encourage OSS maintainers to enable PSFs, e.g., especially those that are togglable by nature, such as private vulnerability reporting, and recommend OSS platforms provide an \textit{Enable best practices} button for PSFs that require no-to-minimal setup and include links to respective PSFs that are user-friendly for OSS maintainers without security backgrounds.
%\diffadd{
Research challenges for enabling such a button include automatically analyzing the context of the OSS project, including its application-specific code and its dependencies. 
Another research challenge of such a button would be conducting a developer study to determine the best manner in which to incorporate developer preferences into the automation of best PSF practices provided by the button.

%\jhl{
Our study further demonstrates a wide variety of OSS maintainers' views as to the extent to which vulnerabilities should be publicized, especially before a fix is provided. 
That spectrum includes a preference for using email due to the more inherent privacy compared to public GitHub issues or having to deal with the usability issues involving PSFs.
%We discourage OSS maintainers from being complacent with users submitting vulnerabilities using public GitHub issues---which is already discouraged~\cite{gh_issues}---and should instead take advantage of the \textit{Private vulnerability reporting} PSF~\cite{gh2023private} regardless of project size, a PSF we recommend OSS platforms enable by default.
Although GitHub discourages public vulnerability reporting~\cite{gh_issues} and, hence, encourages using the private vulnerability reporting PSF~\cite{gh2023private}, our study provides little evidence to back up such a recommendation from OSS maintainers' perspectives.

Nevertheless, preventing premature disclosure of a potentially severe and easily exploitable vulnerability, which may even be particularly difficult to fix, is a potentially strong argument backing GitHub's recommendation.
To that end, feature research can help automatically identify opportunities for beneficial PSF usage.
For example, when a user tries to submit a public issue mentioning keywords like ``vulnerability'' or ``security'', OSS platforms could display a warning and automatically nudge maintainers to encourage enabling lightweight PSFs that require minimal effort to implement.

Other future research opportunities encouraging the use of private vulnerability reporting include producing a convincing means or incentive of recommending using PSF features (e.g., by gamifying the PSFs with badges or achievements) or even enforcing the use of PSFs that OSS platform providers recommend. 
As an example of such enforcement, future research can automatically convert issues submitted to the public that are likely vulnerabilities or may simply be bugs that, as identified through static reachability analysis, are reachable on the project's attack surface. 
Such research would go beyond the coarse-grained identification of vulnerabilities in a GitHub project's dependencies by helping maintainers triage potential vulnerabilities as early as possible 
%(e.g., near bug reporting time) 
with higher confidence (e.g., as provided through a static or dynamic analysis during report submission).
Such a future approach would help alleviate the steep learning curve of understanding software security
%---which is often not the expertise of an OSS maintainer---
while promoting the adoption of PSF features that can reduce exploitation of publicly-reported OSS vulnerabilities.
%}

OSS platforms and researchers should also consider working towards functionalities that (1) improve the quality and presentation of security notifications; (2) automatically inform contributors about if their issue should be filed as a vulnerability report based on past OSS vulnerabilities; %\diffadd{
(3) provide PSF setup assistance, e.g., generating a context-aware security policy;%}
%or configuring static code analysis tools based on the project;} 
and (4) %\diffadd{
gamification %} 
features to reward projects with recommended security posture and be designed in such a way that creates a positive reputation.
Research in such directions can focus on creating better-designed PSFs to provide more value to OSS maintainers, such that PSFs are easier to use, provide guidance to reporters, and promote positivity with their adoption.
%allow access control to other collaborators to view and contribute to vulnerability fixes without administrative privileges.
%Considering these goals in mind, OSS platforms can work towards creating usable PSFs for OSS maintainers without security backgrounds.
%\diffadd{
Future research can also investigate how to effectively use gamification to disseminate user-friendly resources on proper OSS security, allowing any maintainers to effortlessly learn security best practices. For example, a project reputation icon, e.g., ``green shield'', for enabling best-practices PSFs can come with documentation on why each PSF should be enabled. % when the icon is hovered over.
Further, PSFs can have a leaderboard of many kinds: maintainers who followed security best practices most, reporters who provide the best-rated (e.g., by maintainers) vulnerability-oriented issues, reporters who provide the easiest-to-understand and most convincing fixes (e.g., avoiding regressions while ensuring security), and even OSS projects with the best security-focused documentation.
%\diffadd{OSS platforms should also consider security funding for maintainers and reporters. \josh{What's the potential research challenge in terms of this funding? How about more gamification instead as a form of incentive (e.g., building a better reputation for the project, maintainer, or reporter)? For example, a leaderboard of many kinds: most secure projects in terms of a wide variety of security and privacy metrics, maintainers who followed security best practices most, the reporters who provide the best-rated (e.g., by maintainers) GitHub vulnerability-oriented issues, the reporters who provide the easiest-to-understand and most convincing fixes (e.g., in terms of avoiding regressions while ensuring security and privacy), the reporters ranked by their adherence to responsible disclosure practices, and even OSS projects with the best security and privacy-focused documentation.}
This can further motivate maintainers to engage with security-related gamification features and available security resources.%}
\vspace{-0.3cm}

\section{Conclusion}\label{sec:conclusion}

In this paper, we conduct a mixed-methods study, i.e., one listing survey and semi-structured interviews, to investigate challenges and barriers by systematically identifying and quantifying the factors that affect how OSS maintainers, who manage previously vulnerable projects, conduct vulnerability management. 
OSS project maintainers find trusting the supply chain and lack of automation to be the most challenging aspects of vulnerability management.
Further, a lack of awareness, complexity, and perception of being unnecessary are the most prominent barriers to setting up platform security features. 
%We also find that assisted analysis and triaging, informed platform security setup, and the ability to apply for security-specific funding are the most wanted capabilities on OSS platforms.
Based on our findings, future work should investigate secure CI features for vulnerability management, improving PSF usability, and automated vulnerability management tooling. %%%(e.g., vulnerability scoring).    
\section{Ethics considerations}\label{sec:ethics}

Our institution’s ethics review board approved both studies. 
Participants signed consent forms detailing study plans and participant rights before data collection. Further, our study is GDPR-compliant.
We exclude subjects who reside in an OFAC-sanctioned region and/or are affiliated with an OFAC-sanction entity, as required by our institution’s ethics review board for the nature of an international, online study.

Participants were asked to familiarize themselves with consent and data handling information on a study information sheet before agreeing to participate in our studies. Considering the nature of our questions regarding vulnerability-related incidents, in particular to interviewee participants, we state upfront how participants skip any questions, e.g., that they may not be comfortable with answering, or end the interview anytime. Lastly, we shared a preprint of our study to ensure they were okay with our use of their quotes and referencing respective conversations.

%\diffadd{
We did not monetarily compensate survey participants because of (1) our funding limits as academic researchers, and (2) similar to other qualitative studies, compensating international participants is logistically difficult. 
We focused on creating a study that provided intrinsic value to participants through meaningful engagement and insights.
Further, (3) we reduced the time cost of completing our survey by allowing participants to skip any free-response questions and encouraged them not to spend more than 15 minutes filling out our survey, and (4) survey participation is voluntary, OSS participants often support academic research out of intrinsic motivation~\cite{fischer2024chall,gerosa2021shifting,hars2002working} to help improve the security posture of the OSS ecosystem through our study’s results. This approach ensured that participation gives a sense of purpose and contributing to the greater good, i.e., for further securing the OSS ecosystem, despite the lack of monetary compensation for survey participants. However, each of our 22 interviewees was monetarily compensated \$20.
%}

The most recent human-centered security paper that recruited GitHub subjects, published at PETS, mined commit information for maintainer emails~\cite{utz2022privacy}. After consulting with ethics reviewers, they concluded that moving forward, researchers should \textit{``only use contact information that has visibly been made public by the individuals themselves with the intention of allowing the general public to contact them [since] GitHub’s email address mechanics and users’ lack of knowledge about them had neither been mentioned nor addressed by previous work that used public GitHub repositories for recruitment''}~\cite{utz2022privacy}. To comply with GitHub's terms of service~\cite{githubTOS} and following the PETS paper statement, we only reached out to maintainers who have publicly available contact information advertised as reachable to the general public, e.g., in text as a part of their profile introduction markdown, or through a self-hosted website, e.g., a personal homepage.

%\vspace{-0.25cm}

%\diffadd{
\section{Open science}

\noindent We make our listing study survey questions, semi-structured interview guide, detailed interview participants' information, and an extended table of codes available in our artifact~\cite{icse25repo}. 
%}

\vspace{0.2cm}

%Upon acceptance, we will convert our Appendices to a more comprehensive artifact and host it on a licensed platform, e.g., Zenodo, to comply with page limits as required by USENIX Security.
\bibliographystyle{plain}
\bibliography{main}
%\input{content/appendix}

%%%%%%%%%%%%%%%%%%%%%%%%%%%%%%%%%%%%%%%%%%%%%%%%%%%%%%%%%%%%%%%%%%%%%%%%%%%%%%%%
\end{document}